\documentclass[10pt,times]{article}

\usepackage[margin=1in]{geometry}
\usepackage{graphicx}
\usepackage[dvipsnames]{xcolor}
\usepackage{color}
\usepackage{comment}
\usepackage{makecell}
\usepackage{booktabs}
\usepackage{wrapfig}
\usepackage{multirow}
\usepackage{xspace}
\long\def\symbolfootnote[#1]#2{\begingroup%
\def\thefootnote{\fnsymbol{footnote}}\footnote[#1]{#2}\endgroup}
\usepackage{amsmath} 
\usepackage{amsthm}
\usepackage{amsfonts} 
\usepackage{natbib}

\usepackage{hyperref}
\usepackage{float} 
\restylefloat{table}
\usepackage{tabularx}
\usepackage{microtype} 
\usepackage{subcaption}
\usepackage{enumitem}
\usepackage{dirtytalk}
\usepackage{url} 
\usepackage[title]{appendix}

\graphicspath{{figures_reduced/}}

\def\firstmarkupcolor{black}
\def\secondmarkupcolor{black}

\newcommand{\firstrevision}[1]{\textcolor{\firstmarkupcolor}{#1}}
\newcommand{\secondrevision}[1]{\textcolor{\secondmarkupcolor}{#1}}

\usepackage{authblk}
\makeatletter
\renewcommand\AB@affilsepx{, \protect\Affilfont}
\makeatother

\usepackage{fancyhdr}
\begin{document}

\title{THERIF: A Pipeline for Generating Themes for Readability with Iterative Feedback}

\author[1]{Tianyuan Cai}
\author[2]{Aleena Gertrudes Niklaus}
\author[2]{Michael Kraley}
\author[2]{\\Bernard Kerr}
\author[1]{Zoya Bylinskii}
\affil[1]{Adobe Research}
\affil[2]{Adobe Inc.}

\maketitle

\begin{abstract}
Digital reading applications give readers the ability to customize fonts, sizes, and spacings, all of which have been shown to \firstrevision{improve the reading experience for readers from different demographics}. However, tweaking these text features can be challenging, especially given their interactions on the final look and feel of the text. Our solution is to offer readers preset combinations of font, character, word and line spacing, which we bundle together into reading themes. To arrive at a recommended set of reading themes, we present our THERIF \secondrevision{pipeline}, which combines crowdsourced text adjustments, ML-driven clustering of text formats, and design sessions. We show that after four iterations of our \secondrevision{pipeline}, we converge on a set of three COR themes (Compact, Open, and Relaxed) that meet \secondrevision{diverse readers' preferences}, when evaluating the reading speeds, comprehension scores, and preferences of hundreds of readers with and without dyslexia, using crowdsourced experiments.
\end{abstract}

\section{Introduction}

From the moment we wake up to the moment we put down our personal devices at night, we consume most of our information in digital form: news and social media on mobile devices, work emails and documents on our computers, and leisurely reading on our e-readers. Increasingly, applications we use for reading are tailored to these devices and our preferences. Amazon's Kindle allows for text setting adjustments like font size and screen contrast; Microsoft's Immersive Reader increases the accessibility of the text through increased character spacing and text-to-speech options; Adobe Acrobat's Liquid Mode hands control of font size, line, and character spacing to the reader. Customization and accessibility go hand in hand, as readers increasingly gain control of the format in which they consume information.

Findings and best practices from education, design, user interface, and human vision communities point to features of the text --- like serifs, particular stroke widths, font sizes, and spacings --- that can benefit readers with dyslexia, readers of old age, children learning to read, etc.~\citep{rello2012layout, Franken2015, Dogusoy2016, Beymer2008, li2020controlling, Bernard2001, hanson2005personalization, Banerjee2011a, Tai2012}. At the same time, a body of literature is emerging to demonstrate that large reading gains are possible by individuating font and other text characteristics to each reader, young or old, proficient or struggling~\citep{Calabrese2016, Chatrangsan2019, Smither1994, Rello2017, cai2022personalized, wallace2022towards, Banerjee2011, Beier2009, Rello2016, rello2013good, Beier2013, Beymer2008, Bernard2002, Bernard2003, Bhatia2011, Boyarski1998, Poulton1965, Wilkins2009}. However, aside from increasing text size to be more legible, readers may not know which text settings may affect their reading the most. Moreover, many of the features are interrelated, where adjustments to character spacing, for example, may require further adjustments to word or line spacing to feel comfortable. This is a difficult text formatting problem to leave in the hands of casual readers.

\begin{figure}
      \includegraphics[width=\textwidth]{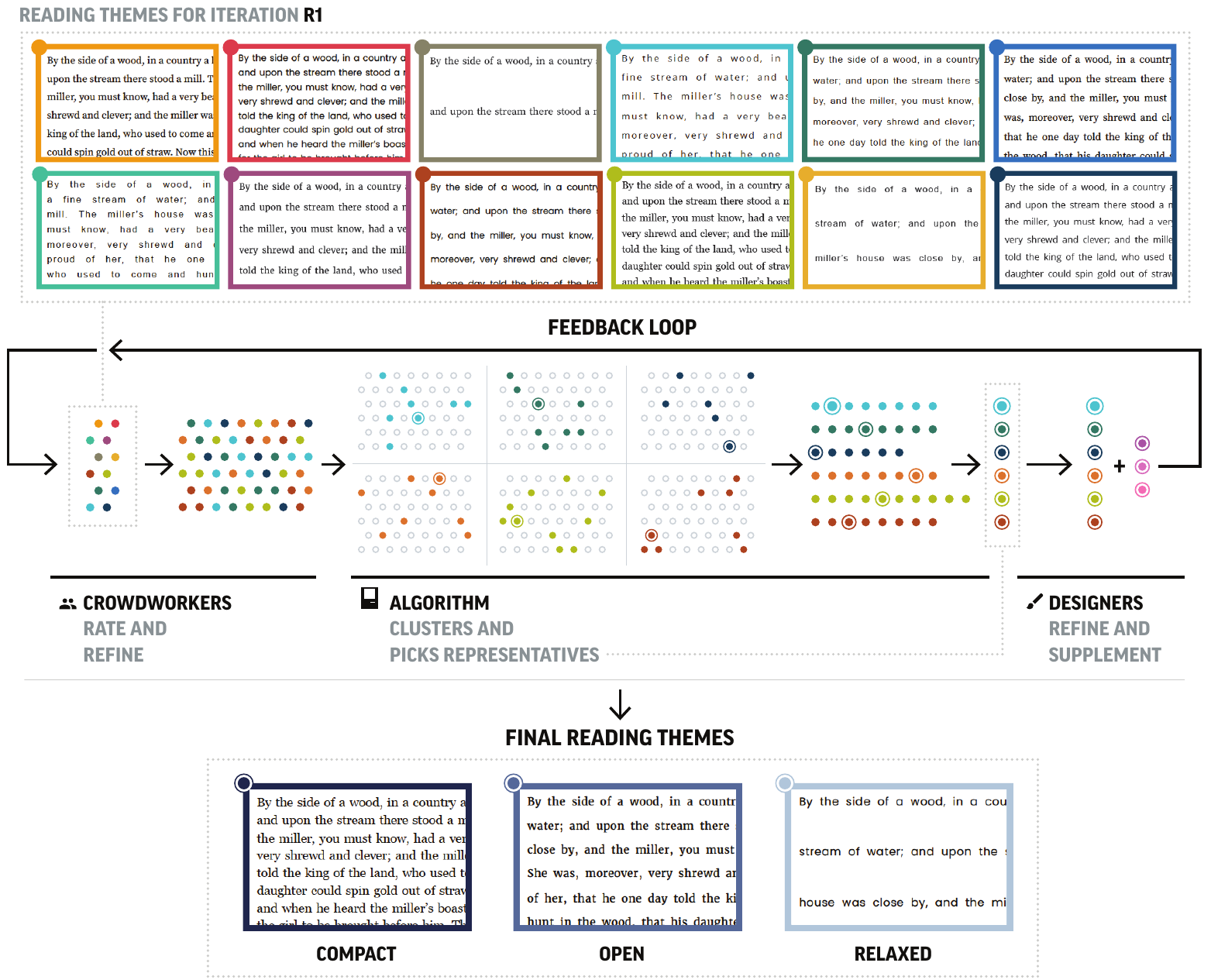}
      \caption{We present a \secondrevision{pipeline} that generates \textbf{the}mes for \textbf{r}eadability with \textbf{i}terative \textbf{f}eedback (THERIF). Each iteration is initialized with text presets bundled into reading themes that crowdworkers rate and refine. The resulting personalized text formats are then automatically clustered using an ML-based algorithm, and cluster representatives are selected to serve as themes for the next iteration. Designers refine and supplement the automatically generated themes with additional themes before the iterative loop continues with a new set of crowdworkers. Each colored dot represents a unique text format, with variations to font, as well as character, word, and line spacing. Some crops from example text formats produced in a single iteration of THERIF are displayed at the top. At the bottom are the three \firstrevision{COR (Compact, Open, Relaxed)} reading themes obtained after four iterations of THERIF.}
      \label{fig:teaser}
\end{figure}

To close this gap and bring readers closer to text formats that are best for them, we bundle fonts and spacings together to offer readers starting points for their custom reading formats that we call \textbf{reading themes}. Recognizing that we want the reading themes to both fit diverse readers' preferences and be well-designed for future use in reading applications, we \firstrevision{follow established approaches in crowdsourcing design and inclusive design guidelines to} introduce a \secondrevision{pipeline} for generating \textbf{the}mes for \textbf{r}eadability with \textbf{i}terative \textbf{f}eedback, that we refer to as \textbf{THERIF}. In this \secondrevision{pipeline}, we continually iterate through crowdsourced text setting refinements, automatic clustering, and design sessions (Figure~\ref{fig:teaser}). With a focus on English reading, we show that each such iteration improves the reading themes, which become more representative of \secondrevision{diverse} reader preferences, require fewer refinements, and are perceived as more likable by readers. After four iterations, we converge on three reading themes that can be deployed in reading applications. \firstrevision{We also show that the THERIF-generated themes can offer} improvements to comfort, comprehension, and speed compared to baseline reading experiences.

Our main contributions include: (1) an open-source prototype to customize text formats (available at therif.netlify.app), (2) the THERIF \secondrevision{pipeline} for generating reading themes through multiple iterations of crowdsourcing, automatic clustering, and design sessions; and (3) a proposed set of three COR (Compact, Open, Relaxed) reading themes representative of \secondrevision{diverse} reader preferences \firstrevision{(CSS available in Appendix~\ref{appendix:final-themes-css})}.

\section{Related Work}

\subsection{A multitude of factors influence digital reading}
\label{sec:relworkadjustments}

A variety of text settings affect individuals' comfort and performance when reading digitally. Font can significantly affect readability~\citep{Banerjee2011,Beier2009,Rello2016,rello2013good,Beier2013,Beymer2008,Bernard2002,Bernard2003,Bhatia2011,Boyarski1998,Poulton1965,Wilkins2009}, which may be attributed to characteristics such as font weight~\citep{oderkerk2020fonts}, stroke contrast~\citep{beier2021high, beier2021increased, beier2021readability}, and character width~\citep{minakata2021effect, ohnishi2021effect}. Serif and sans serif fonts do not differ significantly in legibility~\citep{MohamadAli2013, arditi2005serifs}, but the increase in spacing due to the inclusion of serifs has been shown to have a positive effect on reading~\citep{arditi2005serifs}.  Importantly, font characteristics affect different individuals differently, and there is no one-size-fits-all font~\citep{Calabrese2016, cai2022personalized, wallace2022towards}.

Spacings affect digital reading, and their effects similarly vary by reader. Larger character spacing may benefit readers with dyslexia~\citep{Rello2017, marinus2016special, zorzi2012extra}, with low vision~\citep{beier2021increased}, and those reading unfamiliar content~\citep{Tai2012}. \cite{Rello2017} found larger character spacing to also benefit readers without dyslexia, while \cite{korinth2020wider} found it to hinder proficient readers. Compared to character spacing, word spacing has received less attention. \cite{reynolds2004you} studied children learning to read and found larger word spacing to produce positive reading outcomes. Past work recommended increasing word spacing proportionally with character spacing to avoid compromising reading performance~\citep{reynolds2004you, galliussi2020inter}. To our knowledge, no consistent recommendation for line spacing exists, but previous work and guidelines often recommended maintaining at least a single line spacing~\citep{Rello2016, kirkpatrick2018web, dyson2004physical}. Other text formatting factors including text alignment, paragraph indent, and paragraph spacing have been shown by prior work to have little effect on reading performance~\citep{miniukovich2017design, british2012dyslexia, barrow2010make, rainger2003dyslexic, aziz2012interaction, de2012web}.

\subsection{Personalizing text settings is challenging}
\label{sec:personalizing-text-settings-is-challenging}

Many recommendations exist for tailoring text settings to individuals (\S\ref{sec:relworkadjustments}), but adopting them can be challenging for casual readers. Reader characteristics such as dyslexia and language fluency vary on a continuous spectrum~\citep{cooper2011revised, snowling2012validity}, but recommendations for text settings are largely made based on demographic categories (e.g., \secondrevision{dyslexia fonts, child-friendly formats}, etc.), without consideration for overlap in reader needs and reader characteristics that vary on a continuum. Additionally, recommendations often do not account for relevant contexts, such as time of day, type of reading, or situationally-induced impairments and disabilities (SIIDs)~\citep{darroch2005effect, keenan2008reading, yamabe2007experiments}. For instance, \cite{yamabe2007experiments} found that individuals reading on their mobile devices while walking may benefit from larger font sizes, similar to low-vision readers.  Furthermore, existing reading studies often focus on objective measures such as reading comprehension and speed~\citep{cai2022personalized, wallace2022towards, Chatrangsan2019, zhu2021effects, Rello2016, McKoon2016}. However, factors such as comfort and personal preference are equally important to readers when reading digitally, although consistent criteria may not always exist~\citep{zhu2021effects, Bernard2003, Kulahcioglu2019}. Recommending text settings while balancing multiple objectives and meeting diverse readers' \secondrevision{preferences} is a challenging problem, which may explain why there is limited work beyond personalized font recommendation~\citep{cai2022personalized}.

Guiding readers but leaving them with agency over the final reading formats may offer greater flexibility to their local context~\citep{bentley1995medium}. However, the reading and text setting interfaces available today are not well suited for this purpose, as continuous adjustments to spacing and long drop-down menus of fonts are challenging for casual readers to navigate. \cite{odonovan2014} highlighted that the font selector is often alphabetically ordered with limited guidance on which font works the best. \cite{hanson2005personalization} found that when offered the ability to adjust spacing and zoom levels, participants were often unfamiliar with these settings and reluctant to make changes without explicitly understanding how they may affect reading formats. These problems point to the need to guide readers towards presets, and give them good defaults, or starting points for further customization.

\subsection{Clustering into presets can simplify the space of options}

Previous literature proposed ways to adapt interfaces to \secondrevision{individual needs}, but these recommendations were often inflexible to the varied circumstances readers find themselves in~\citep{darroch2005effect, keenan2008reading, yamabe2007experiments, cooper2011revised, snowling2012validity}. On the other hand, providing too many text customization options may be overwhelming for casual readers (\S\ref{sec:personalizing-text-settings-is-challenging}). One solution is to bundle settings into several \emph{presets} for users to select from, a common approach in similar situations~\citep{nebeling2021xrstudio, ackerman2005privacy, grudin2004managerial, olson2004toward}. For instance, \cite{nebeling2021xrstudio} found that the creation of video presets allows video editors to conveniently export footage to a suitable platform without fine-tuning audio, video, and caption settings individually. By offering multiple presets, designers can tailor the same interface to groups of users with different preferences. \cite{grudin2004managerial} identified the need to offer different presets of software configurations based on the user's job functions. In the absence of explicit user grouping, researchers clustered users with similar characteristics and presented them with tailored interfaces to help them more easily configure complicated settings~\citep{ackerman2005privacy, olson2004toward}. Leveraging these learnings, we developed presets of text settings and assessed their effectiveness for reading. We refer to these presets as ``reading themes''.

Clustering user preferences allows for the development of interface experiences that meet \secondrevision{diverse preferences}~\citep{pruitt2003personas}. Researchers have experimented with clustering approaches with and without human intervention. While unsupervised learning algorithms can automatically cluster similar user preferences to facilitate interface design~\citep{salminen2020literature, guan2016apparel, gasparetti2007exploiting}, frequently, collaborations between experimenters, experts, and unsupervised algorithms are necessary when clustering based on unstructured information, such as audio and visual data~\citep{chuang2012interpretation}. On the other end of the spectrum, \cite{pruitt2003personas}, for instance, used an experimenter-driven approach that clustered qualitative and quantitative evidence to identify groups of user preferences.

Recruiting human expertise may be time-consuming and unaccommodating to large data sets~\citep{chilton2013cascade, chang2016alloy}, but it may be necessary when a semantic understanding of the user experience is important. There is a variety of approaches to efficiently involve experts when clustering. For instance, \cite{preston2010redefining} invited experts to construct a matrix of constraints to facilitate cluster convergence. \cite{awasthi2014local} involved users to improve existing cluster partitions by performing ``merge-and-split''. In this work, we considered expert designer feedback to supplement our automatic clustering procedure, but eventually found that our automatic clustering was able to achieve comparable results.

\subsection{\protect\firstrevision{Crowdsourced design processes capture user preferences}}
\label{sec:relwork-crowdsourcing}

\secondrevision{ Individual readers may struggle to converge on optimal reading formats due the complexity of existing text setting interfaces~(\S\ref{sec:personalizing-text-settings-is-challenging}). An entirely designer-driven process has the risk of failing to represent varied reading needs~(\S\ref{sec:relworkadjustments})~\citep{bennett2019promise}. Therefore, we explore ways to combine designer and participant input by drawing learnings from prior works on crowdsourcing design, expert participation, and scalable feedback.}

\secondrevision{\paragraph{Crowdsourcing design} Previous works explored large-scale user participation in individual design steps~\citep{salganik2015wiki, cranshaw2011polymath, xu2012you}, and in the entire design process~\citep{yu2011cooks, park2013crowd}. Crowdworkers have shown that they can collectively create designs with high originality and quality despite having little design expertise~\citep{chen2013comic, yu2011cooks, nickerson2008spatial}. For instance, \cite{yu2011cooks} showed that workers from Amazon Mechanical Turk could construct creatively designed chairs. \cite{komarov2013crowdsourcing} found that crowdsourcing helps achieve similar results as in-lab participation while facilitating greater participant diversity, an important consideration in our study due to the importance of individuated reading~(\S\ref{sec:relworkadjustments}).}

\secondrevision{\paragraph{Designer participation} Although researchers have crowdsourced effective designs in the absence of designers~\citep{yu2011cooks}, design input can improve design quality~\citep{spinuzzi2005methodology}. In the design of reading formats, such input can ensure alignment with typographical guidelines~(\S\ref{sec:personalizing-text-settings-is-challenging}). 
Additionally, involving designers may help break ties when multiple designs emerge as possibilities~\citep{briggs2003collaboration, merz2016exploring, park2013crowd}. For instance, \cite{park2013crowd} paired teams of crowdworkers with designers, and found that involving designers helped encourage exploration of diverse ideas and convergence on final crowdsourced designs which were rated highly by external experts. However, involving expert participation may be infeasible for reviewing crowdsourced designs at scale~\citep{park2013crowd,head2017writing,glassman2015overcode}.}

\secondrevision{\paragraph{Scalable feedback} To elicit expert feedback on large scale submissions, previous work explored submission clustering so that experts can contribute their knowledge more efficiently~\citep{glassman2015foobaz, glassman2015overcode,moghadam2015autostyle,head2017writing}. For instance, \cite{head2017writing} used program synthesis to cluster programming code based on the similarity of the underlying issue such that instructors could provide feedback about a cluster (or its representative) rather than doing it for each submission. We leverage a similar approach but utilize a convolutional neural network when clustering the crowdsourced designs of reading formats, since it allows perception-based clustering. By clustering crowdsourced designs for designer feedback, we explore ways to balance crowdsourcing designs at scale with the inclusion of designers' expertise in the creation process~(\S\ref{sec:therif}). Additionally, we repeat the design process iteratively since multiple design iterations have been shown to lead to improved design solutions over time in crowdsourcing design setting~\citep{gulley2001patterns, resnick2009scratch, yu2011cooks, xu2015classroom}}.

\secondrevision{These three components - a combination of crowdsourcing, design iterations, and automated clustering for scalability - form the foundations of our THERIF pipeline for generating reading themes.}

\section{\protect\firstrevision{Eliciting reading preferences}}\label{sec:pilot}

Different reading applications offer control over different text settings such as font choice and size, character, word and line spacing. At the same time, prior work has shown that tuning the text format to the individual reader can significantly improve reading performance (\S\ref{sec:relworkadjustments}). However, the space of possible text adjustment settings is large, and systematically iterating over combinations of these text features would be intractable. \firstrevision{We leverage the volume and diversity of crowdworkers to sample this space based on their preference~(\S\ref{sec:relwork-crowdsourcing}).} For this purpose, we built a prototype that offers readers fine-grained control over their text settings, and put this \textbf{text settings prototype} in front of crowdworkers to discover which settings are most commonly adjusted and used together \firstrevision{to arrive at their preferred reading formats}.
In this section we describe our prototype and how its design evolved with learnings from the pilot study. The final prototype design was used in our THERIF \secondrevision{pipeline} for generating reading themes (\S\ref{sec:therif}). Data from the pilot study were also used to initialize the reading themes in THERIF (\S\ref{sec:initialize-therif}).

\subsection{\firstrevision{Text settings} pilot study}\label{sec:readingcontrols}

\begin{figure*}[!htbp]
      \centering
      \begin{subfigure}{.45\textwidth}
            \centering
            \includegraphics[width=\linewidth]{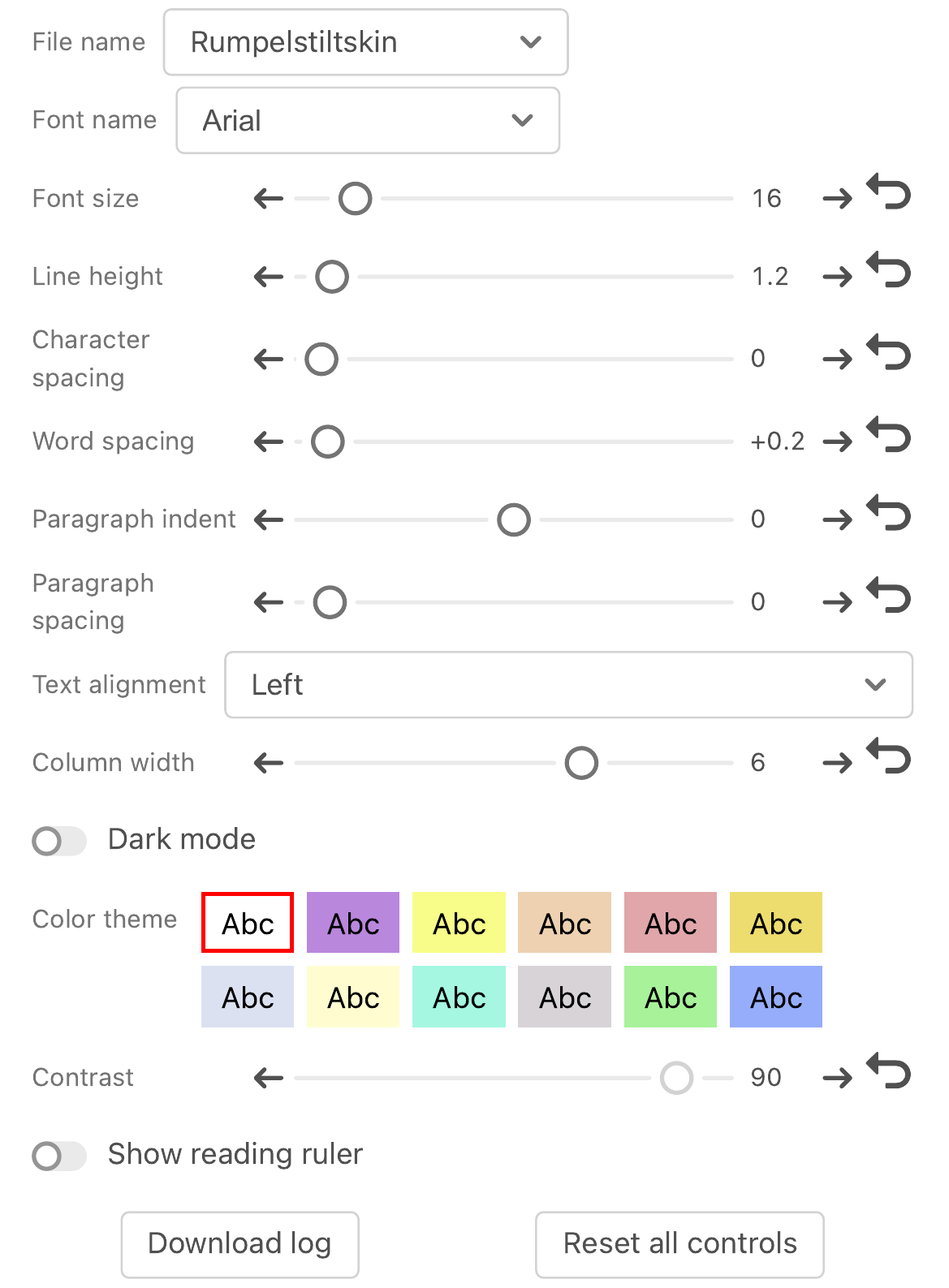}
            \caption{Text Settings Panel (pilot study)}
            \label{fig:study-interface-pilot-text-settings}
      \end{subfigure}
      \begin{subfigure}{.45\textwidth}
            \centering
            \includegraphics[width=\linewidth]{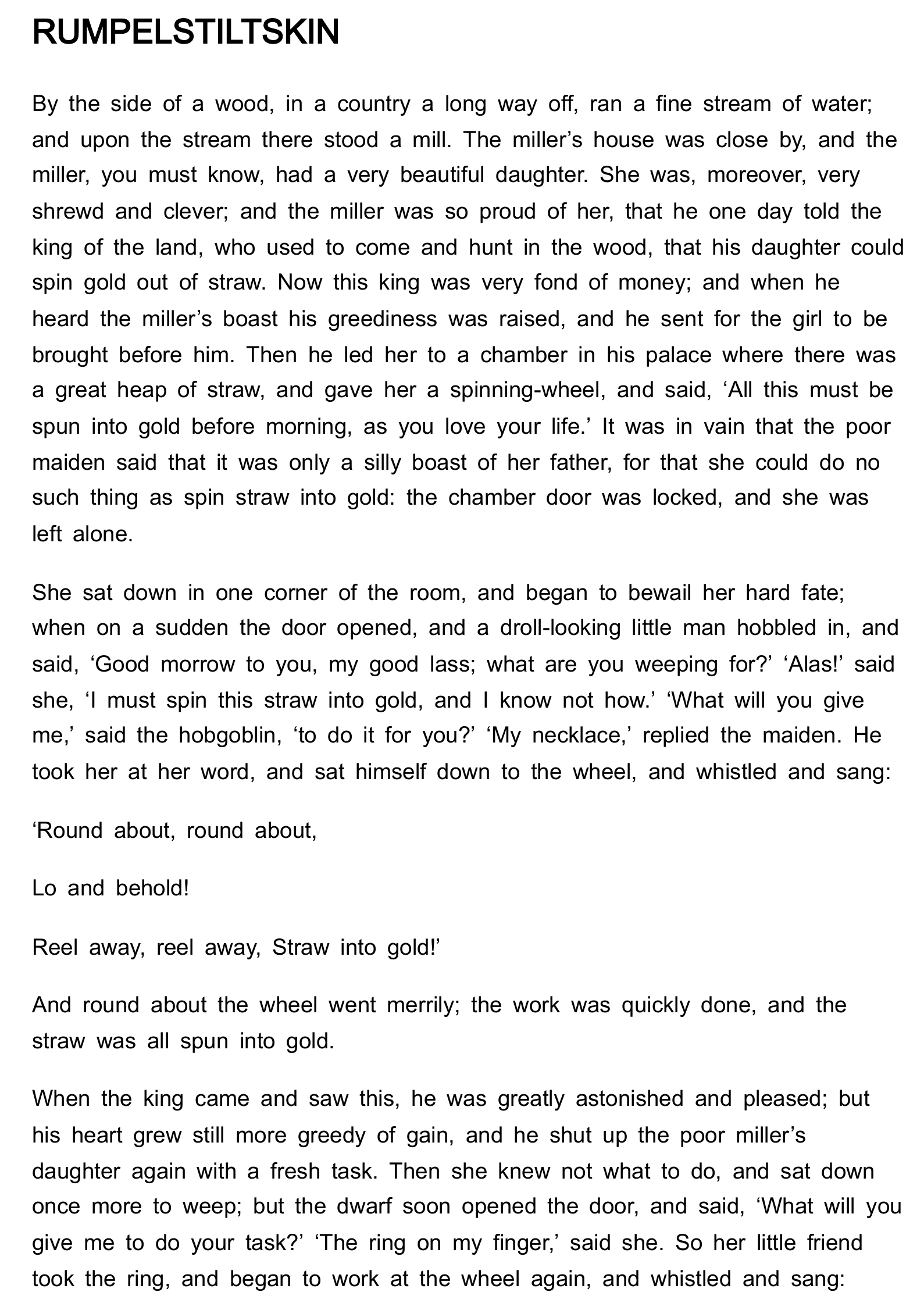}
            \caption{Reading Panel (pilot \& main studies)}
            \label{fig:study-interface-pilot-reading-preview}
      \end{subfigure}
      \caption{  The left and right panels of the study interface used during the pilot study. The left panel (a) offers control over a comprehensive set of text settings, and the right panel (b) shows a preview of the reading format based on the current settings. Participants can view the effect of their text settings on four different passages. The ``reading ruler'' option was not used for this paper, and participants in the pilot study were not instructed to use it.}
      \label{fig:study-interface-pilot}
\end{figure*}

The first version of our text settings prototype offered participants an exhaustive list of 11 possible text adjustments based on prior literature (\S\ref{sec:relworkadjustments})~\citep{miniukovich2017design, miniukovich2019guideline, hanson2005personalization}, including fonts, sizes, spacings, text alignments, and color themes, among others (Figure~\ref{fig:study-interface-pilot}). We offered eight fonts: Montserrat, Open Sans, Arial, Roboto, Merriweather, Georgia, Source Serif Pro, and Times, based on previous literature showing that they are diverse, prevalent, and readable, with characteristics generalizable to other fonts~\citep{cai2022personalized}. Participants could adjust these settings in the \textbf{text settings panel} (Figure~\ref{fig:study-interface-pilot-text-settings}) and preview the changes to the text in a \textbf{reading panel} (Figure~\ref{fig:study-interface-pilot-reading-preview}) preloaded with four Creative Commons passages in English (723-2934 words each). All text settings included wide ranges of possible values and supported adjustments in both step increments and with a continuous slider. In later parts of our study, a \textbf{theme review panel} also allowed participants to review and rate text setting presets (Figure~\ref{fig:study-interface-user-rating}).

We conducted a pilot study with the prototype to assess its ability to support the THERIF \secondrevision{pipeline} (introduced in \S\ref{sec:therif}). The study included four iterations and was conducted on the UserTesting platform~\citep{usertesting}, using the think-aloud protocol to collect participants' feedback. The study involved 271 crowdworkers and design sessions with four designers throughout four iterations. Iterations included 100, 51, 60, and 60 crowdworkers, respectively. Studies lasted an average of 40 minutes, with each participant compensated \$30 hourly.

\firstrevision{The four designers\footnote{\firstrevision{Designers were similar to the population they design for~\citep{newel2000user, wobbrock2011ability}. Based on their responses to the dyslexia questionnaire administered, we found that two designers had above-average chances of having dyslexia.}} were recruited from the same large U.S. corporation:
      D1: Man with 27 years of design experience; D2: Man with 20 years of design and typography experience; D3: Woman with 10 years of design experience; D4: Woman with 8 years of design experience. Only full-time work experience was reported.}

In each iteration, crowdworkers used the text settings panel to design their preferred reading formats, \secondrevision{machine learning clustered reading formats into groups containing similar formats, and the designers selected group representatives as the reading themes} for the next iteration~\firstrevision{(\S\ref{sec:NN})}. \firstrevision{Two authors used descriptive codes~\citep{saldana2021coding} to summarize participants' design process and their think-aloud feedback, identifying frequently used text settings and adjustment patterns. The authors then met with all four designers to review the descriptive codes, relate them to participants' designs, and develop high-level learnings~(\S\ref{sec:pilot-initial-learnings}).} We incorporated learnings from the pilot to refine both the prototype and study design for the implementation of THERIF in the next section (\S\ref{sec:therif}).

\subsection{Initial learnings}
\label{sec:pilot-initial-learnings}

\subsubsection{Fewer text settings}

While the first version of our prototype included a comprehensive set of text settings (Figure~\ref{fig:study-interface-pilot}), the pilot study helped us narrow down the text settings that would form the foundation of our reading themes in the following sections. We removed the setting for paragraph spacing and indent because neither designers interviewed in the pilot study nor previous literature considered them important for reading performance. We also removed text alignment because almost all participants chose left alignment, a default supported by typographers and past work~\citep{miniukovich2019guideline, ling2007influence}. We removed settings for background color, contrast, dark mode, and column width due to a lack of consistent patterns in participants' preferences: these settings tended to be dependent on reading context and content, and we imagine that reading applications would add them as customization features alongside the reading themes. The revised prototype is therefore limited to \textbf{font selections, character, word and line spacing \firstrevision{--- the key properties of our reading themes}}. These settings are also those identified to affect web readability by WCAG 2.1~\citep{kirkpatrick2018web}. During discussion, designers similarly mentioned that the combination of glyph and spacing characteristics helps tailor messages to different audiences, and that striking a balance among these variables is challenging. \firstrevision{Appendix} Table~\ref{tab:reading-control-comparison} lists the eleven settings originally included and the four that remained after the pilot study.

\subsubsection{Normalize font sizes}
\label{sec:norm-fontsize}

The viewing distance, screen size, and resolution that participants use for reading (and for this study) are all variable and confounded with the optimal font size~\citep{li2020controlling}. Similarly, when reviewing the reading formats designed by crowdworkers in the pilot study, designers attributed variations in font size settings to device and environment-specific idiosyncrasies. Therefore, we opted to fix the font size in the final prototype.

\begin{figure}[!htbp]
      \includegraphics[width=\columnwidth]{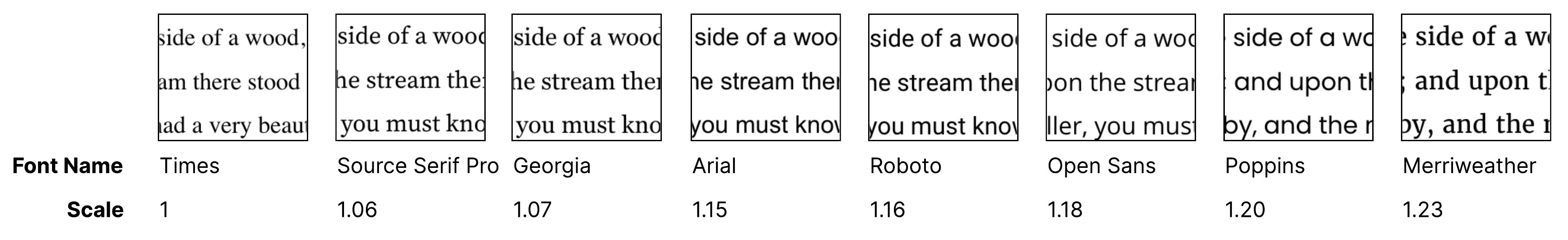}
      \caption{Study fonts vary in perceptual size even when shown at the same pixel size and spacing settings. From left to right, fonts increase in height and decrease in line spacing, and the tallest font is 23\% taller than the shortest. To mitigate bias introduced by unequal font sizes and ensure text settings independently affect reading format, we normalize fonts to have equal x-heights.}
      \label{fig:unnormalized-fonts}
\end{figure}

Even at the same pixel size, fonts with taller x-heights may bias participant preferences~\citep{wallace2022towards}. Designers pointed out that taller fonts in our set, such as Poppins and Merriweather, are up to 23\% taller in x-height than the shortest, and result in a perceptually tighter spacing between lines of text, despite the same spacing setting (Figure~\ref{fig:unnormalized-fonts}). Therefore, similar to previous remote readability studies~\citep{wallace2022towards}, we normalized all fonts to help them appear perceptually similar, and reduce confounds.\footnote{``x-height'' measures the average height of lowercase characters of a font. We performed normalization by x-height rather than glyph height because x-height is one of the key factors that affect the readability of a font~\citep{cai2022personalized}.} In the main study, all fonts have the same x-height as Times at 17px, the most popular font size based on our pilot data.

\subsubsection{Initialize the prototype with more variety}\label{sec:pilotthemes}

In the pilot study, we started all participants with the default setting of Arial font at 16px, 1.2 line spacing, and default values of character and word spacing (0em), and we asked them to explore the text settings to arrive at a preferred format. Designers commented that initializing with the same default setting for everyone may limit the full space of text settings participants end up exploring. On the other hand, starting participants with randomly initialized values for each text setting would lead to many unreadable experiences.  Therefore, we utilized the text setting values selected by participants in the pilot study to initialize the prototype in the main study~(\S\ref{sec:initialize-therif}). Based on clustering the reading formats generated by pilot study participants, we obtained six combinations of \emph{(character spacing {\normalfont (em)}, word spacing {\normalfont (em)}, line spacing)}: (0, 0.2, 1.9), (0, 0.2, 1.6), (0, 0.1, 1.6), (0, 0.1, 1.5), (0, 0, 1.6), (0, 0, 1.5). In the main study, participants in the first iteration of THERIF started with one of these six spacing presets, randomly paired with one of our eight study fonts.  Because previous studies showed diverse preferences for reading fonts~\citep{wallace2022towards}, we randomize the font to ensure all fonts have an equal chance of being chosen.

\subsubsection{Other improvements}

By analyzing participants' think-aloud feedback as they used the text settings prototype in the pilot studies, we made additional improvements to the interface.

Some participants mentioned that the left text settings panel can be distracting when previewing the reading format after adjusting the settings. We updated the prototype design so that the settings panel hides from view when the cursor was moved away, to allow participants to preview the reading passage in isolation of any other UI components, mimicking a naturalistic reading application. In the main study, participants were instructed to move their cursor away from the text settings panel whenever they made adjustments to the text.

We also observed that participants exhibited a higher likelihood of selecting fonts and reading passages positioned towards the top of their respective dropdown lists. To mitigate the positional bias, we randomized the order of fonts and reading passages for each participant in the main study. This final version of the text settings prototype was used in all the iterations of the THERIF \secondrevision{pipeline}, as described in the next section.

\section{Introducing THERIF}
\label{sec:therif}

\begin{figure}[!htbp]
      \includegraphics[width=\textwidth]{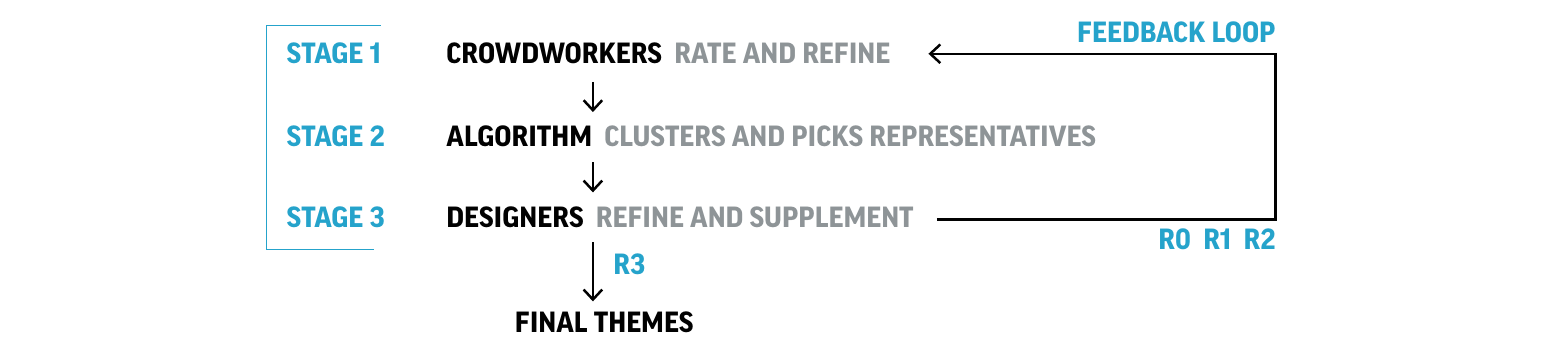}
      \caption{ We introduce a \secondrevision{pipeline} that generates reading themes using an iterative feedback loop composed of three stages: (1)~crowdworkers rate and refine the reading themes from the previous iteration (or from the pilot study in the case of iteration R0) based on their preferences; (2)~ML algorithms automatically cluster the crowdsourced text formats and produce a handful of reading themes for the next iteration; (3)~designers review and supplement additional reading themes. The resulting set of reading themes is shown to a new set of crowdworkers at the start of the next iteration. For our study, we ran four iterations of the feedback loop, though it can be invoked any number of additional times to continue to refine themes. }
      \label{fig:study-design}
\end{figure}

In this section, we introduce the THERIF \secondrevision{pipeline} for producing reading themes. The \secondrevision{pipeline} includes an iterative feedback loop that allows collaboration between crowdworkers, a machine learning algorithm, and designers (Figure~\ref{fig:study-design}). \secondrevision{Components of the THERIF pipeline are motivated by established evidence in crowdsourcing design, designer participation, and scalable feedback systems~(\S\ref{sec:relwork-crowdsourcing}).} 

Each iteration of our THERIF \secondrevision{pipeline} includes three stages: (1) Crowdworkers customize text settings to create diverse reading formats (\S\ref{sec:therif-rate-and-refine}); (2) ML algorithms automatically cluster the resulting formats and designate cluster representatives as \emph{reading themes} for the next iteration (\S\ref{sec:NN}); (3) Designers supplement additional reading themes in order to incorporate any design considerations into the process of theme creation (\S\ref{sec:therif-designer-supplement}). \secondrevision{Stage 1 was motivated by user-centered design and participatory design practices~\citep{spinuzzi2005methodology}, and prior work demonstrating the successful use of crowdsourcing for design studies~\citep{salganik2015wiki, cranshaw2011polymath, xu2012you, yu2011cooks, nickerson2008spatial, park2013crowd, chen2013comic}. Stage 2 leveraged machine learning to simplify collaboration between the crowd and the expert~\citep{head2017writing}. Stage 3 was intended to ensure good design practices were followed in the design of readable text formats~\citep{spinuzzi2005methodology, park2013crowd}. THERIF is run iteratively to help refine designs over time~\citep{gulley2001patterns, resnick2009scratch, yu2011cooks, xu2015classroom}.}

\secondrevision{We repeat these three stages of THERIF over multiple iterations.} We refer to the iterations by R0, R1, R2, and R3. Each iteration starts with a group of crowdworkers adjusting text settings from the provided defaults to their liking. The very first refinement iteration (R0) is initialized with themes that are based on data from the pilot study. All subsequent iterations (R1-R3) are initialized with the themes obtained from the previous iteration, after automatic clustering and design sessions. While we ran four iterations to produce the final set of themes in this paper, this \secondrevision{pipeline} is extensible to future iterations. For every iteration, we recruit a new set of crowdworkers that have not participated in this study before, to ensure that the reading themes evolve to be representative of diverse readers' preferences rather than fine-tuned to the preferences of a few.

\subsection{Stage 1: Crowdworkers rate and refine}
\label{sec:therif-rate-and-refine}

Every iteration of THERIF starts with a new population of crowdworkers using the reading settings prototype (Figure~\ref{fig:study-interface}) to refine the text settings \firstrevision{based on their reading preference}. The prototype is initialized with the set of reading themes from the last iteration of THERIF, except for the very first iteration (R0), which is initialized with a single reading theme for each participant, randomly selected from the pilot study themes. This section describes how we familiarize participants with the text settings and reading themes, and guide them through the process of customizing a desirable reading format.

\begin{figure}[!htbp]
      \includegraphics[width=\columnwidth]{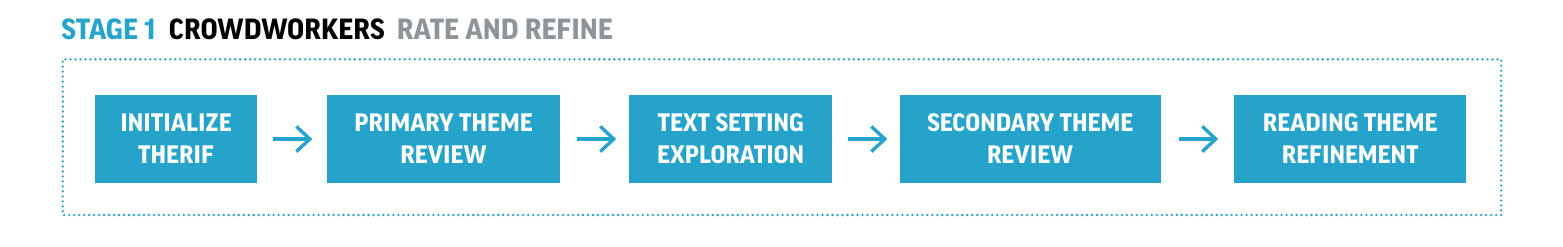}
      \caption{Crowdworkers are guided through a sequence of steps to produce their final text formats. They first explore the available reading themes and text setting adjustments to familiarize themselves with the customization options. Then, they go through a secondary theme review and refinement process to produce the final settings used in our THERIF iterations. }
      \label{fig:stage-one-decomposed}
\end{figure}

\subsubsection{Initializing THERIF}
\label{sec:initialize-therif}

On the very first iteration (R0) of THERIF, our goal was to initialize the reading prototype with some basic text settings that participants could refine further. At the same time, we did not want to bias or anchor participants to any one selection of settings, nor present them with an unreadable default. For this purpose, we used the data from the pilot study to generate six representative combinations of \emph{(character spacing, word spacing, line spacing)} that we sampled from and randomly paired with one of our eight study fonts (\S\ref{sec:pilotthemes}), to generate \secondrevision{a variety of} starting points for participants' text formats.

\subsubsection{Primary theme review}\label{ssec:primaryreview}

\begin{figure*}[!htbp]
      \centering
      \begin{subfigure}{.45\textwidth}
            \centering
            \includegraphics[width=\linewidth]{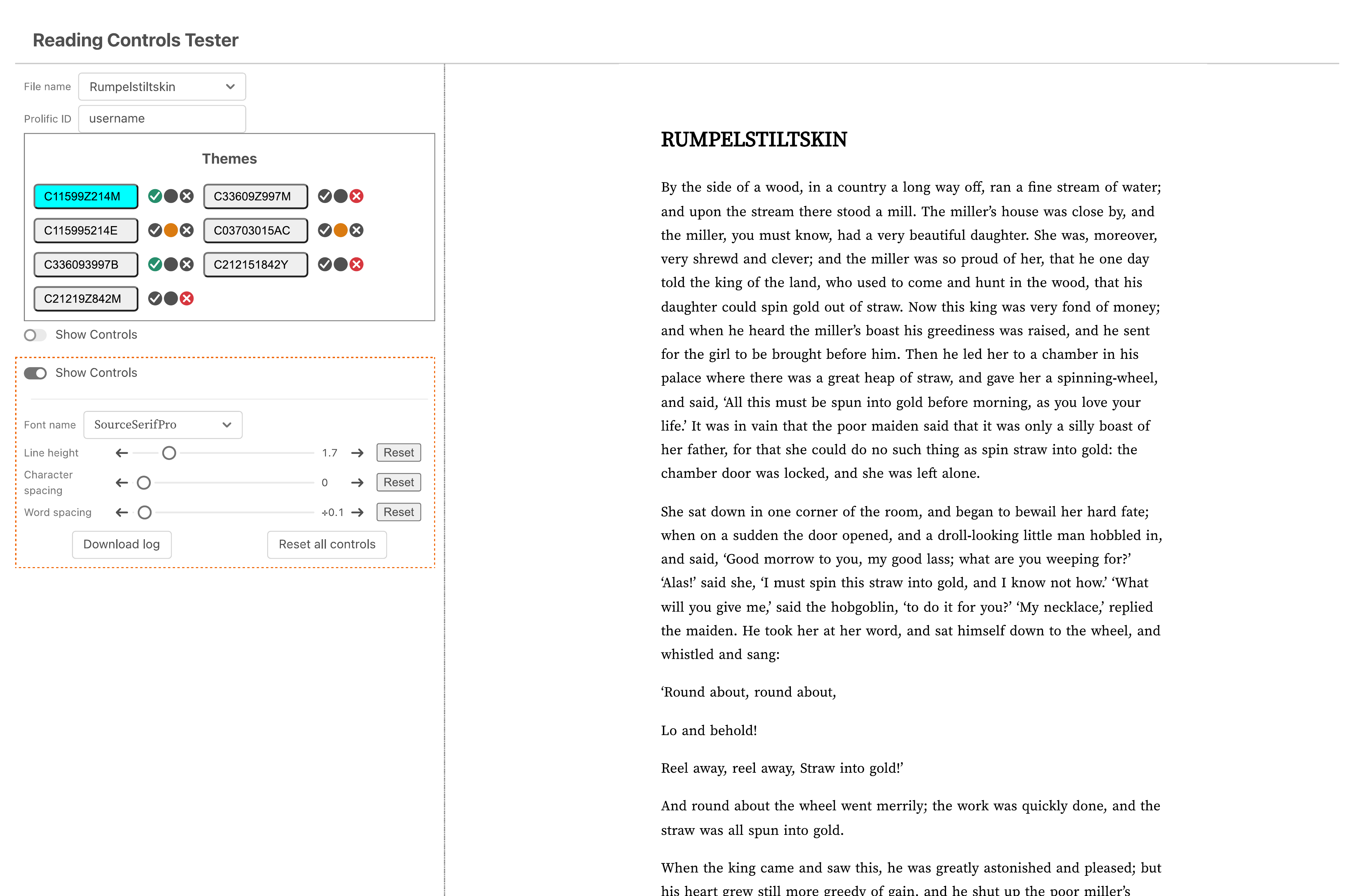}
            \caption{Theme Rating Panel (pilot \& main studies)}
            \label{fig:study-interface-user-rating}
      \end{subfigure}
      \begin{subfigure}{.45\textwidth}
            \centering
            \includegraphics[width=\linewidth]{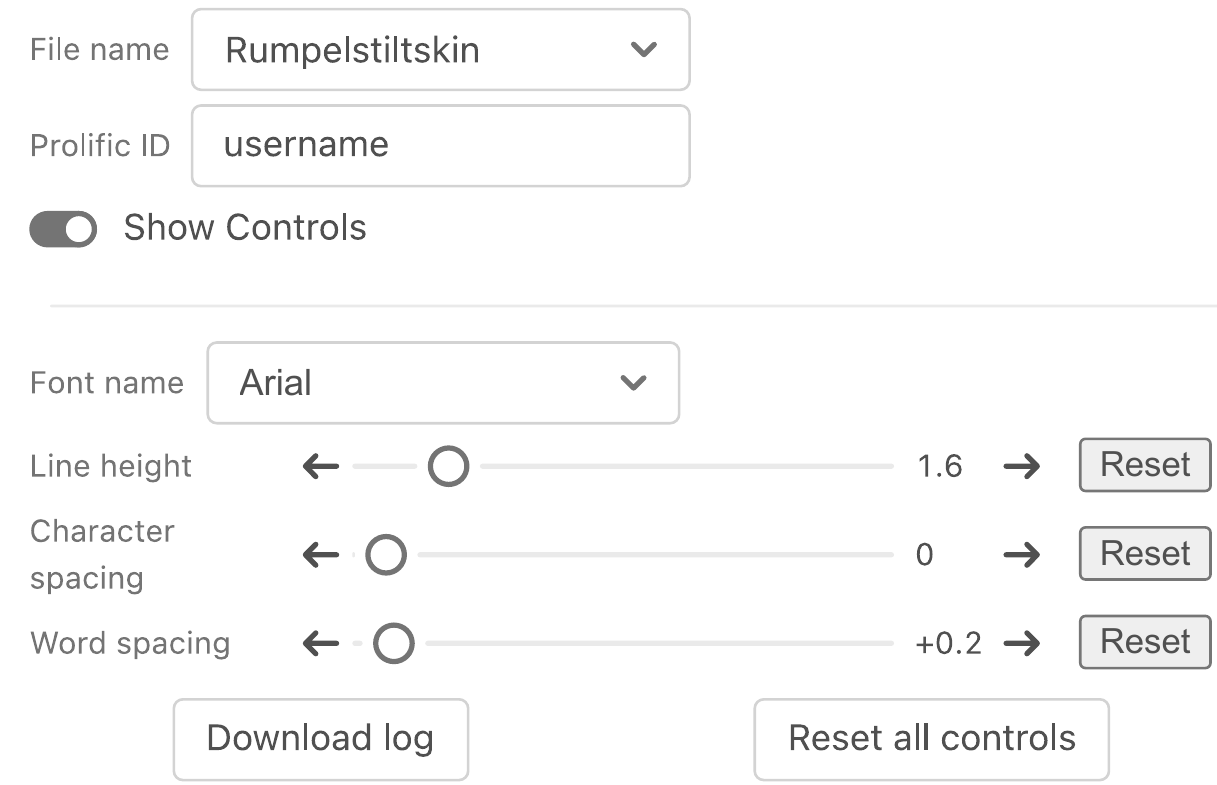}
            \caption{Text Settings Panel (main study)}
            \label{fig:study-interface-toggle-control}
      \end{subfigure}
      \caption{ The left panel used for the main study includes views for theme rating and refinement. Participants first rate each theme (good, unsure, bad) and select their preferred reading theme whenever applicable (Figure~\ref{fig:study-interface-user-rating}). Once finished, they were instructed to toggle on ``Show Control'' to expose the text settings and make further refinements (Figure~\ref{fig:study-interface-toggle-control}). In the main study, the left control panel hides when the mouse moves away, allowing participants better focus on the reading format. \firstrevision{The theme names are anonymized alphanumerical IDs}. See supplementary materials for the full study interface.}
      \label{fig:study-interface}
\end{figure*}

Participants in THERIF iterations R1-R3 first reviewed the reading themes from the previous iteration. They were instructed to review and rate each theme (good, unsure, or bad). We always included a \textbf{validation theme}, intended to represent a poor reading format, in the mix. We selected 11 validation themes from the pilot study design sessions, where we had asked designers to point out poor reading formats from the full set of formats generated by pilot study participants. One of these 11 themes was presented at random along with the other reading themes participants rated. Participants in R0 do not complete the theme rating step because they refine from a randomly initialized reading theme.
Afterward, we asked participants to ``identify your favorite reading theme and click on it''. Their selected reading theme would then be highlighted in the theme rating panel (Figure~\ref{fig:study-interface-user-rating}).

\subsubsection{Text setting exploration}

Starting with their preferred theme (iterations R1-R3) or a randomly initialized theme (R0), participants were then instructed to click a button to uncover the previously hidden text settings for further text refinements. In this part of the study, the theme rating panel (Figure~\ref{fig:study-interface-user-rating}) was replaced by the text settings panel (Figure~\ref{fig:study-interface-toggle-control}). We instructed participants to try out each setting individually and observe how it affected the text using a think-aloud protocol. This ensured that participants gained familiarity with all the text settings before the next step.

\subsubsection{Secondary theme review}

After getting acquainted with the various text settings possible in our reading prototype, we reset the interface. Participants in R0 were now presented with another randomly initialized reading theme, while participants in R1-R3 were once again presented with the theme rating panel (Figure~\ref{fig:study-interface-user-rating}) preloaded with the same themes as before, but in a randomly shuffled order. Participants were asked to rate all the themes once again, and then select their preferred theme. This secondary theme review step allowed participants to make their ratings and selections with awareness of the further text adjustments possible once a theme was selected.

\subsubsection{Reading theme refinement}

Starting with the selected reading theme from the last step, we then asked participants to refine the theme further \firstrevision{``to their liking''}, by toggling to the text settings panel (Figure~\ref{fig:study-interface-toggle-control}) one last time.  As they customized their reading format, we recorded every adjustment, the time elapsed, and the final text settings into a \textbf{refinements log file} for further analysis.

\subsection{Stage 2: Algorithm clusters and picks representatives}\label{sec:NN}

To converge on a handful of reading themes that represent diverse reader preferences, we cluster all the reading formats generated by participants in Stage 1 of THERIF, and \textbf{select cluster representatives to serve as reading themes for the next iteration}. Since hundreds of different readers can arrive at similar formats, reducing this set of formats to representative examples makes evaluation of different format choices more tractable~\firstrevision{\citep{head2017writing}}.

\begin{figure}[!htbp]
      \includegraphics[width=\columnwidth]{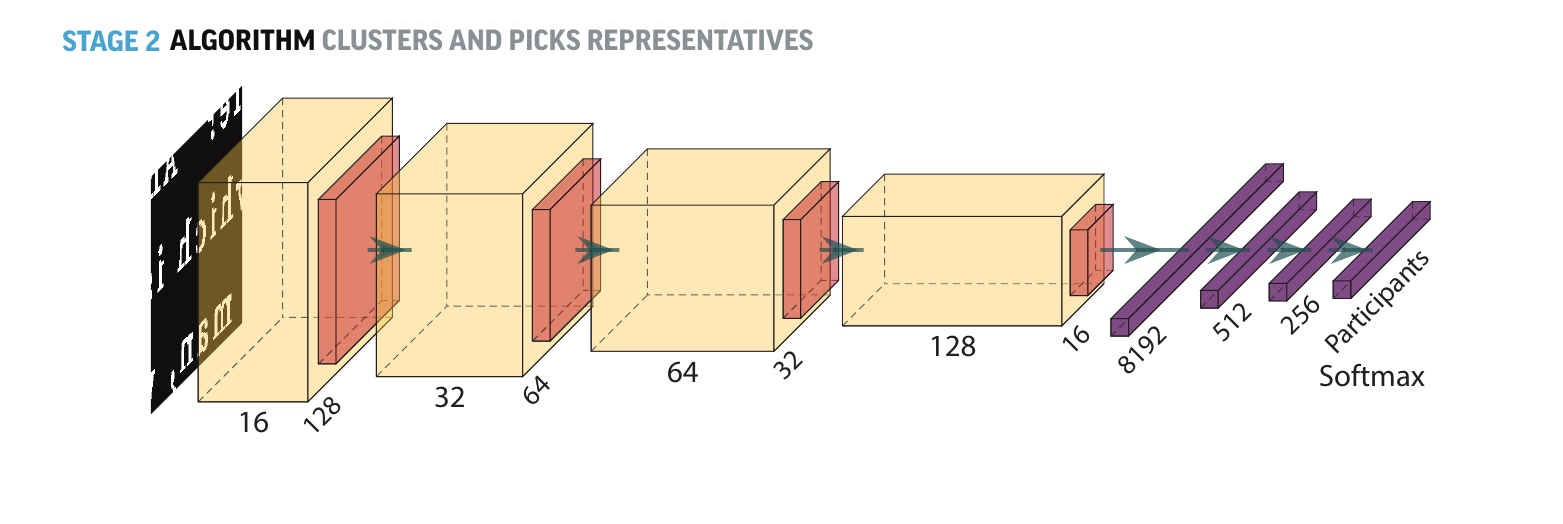}
      \caption{The neural network we trained for encoding reading formats includes several convolutional layers followed by dense layers. Given a crop of text as input, the model learns to predict the participant ID who created the corresponding reading format. This allows the model to learn to encode crops from similar formats similarly. We then use k-Means algorithm to automatically cluster similar reading formats together based on encoded features from this trained model.}
      \label{fig:model-structure}
\end{figure}

\begin{figure}[!htbp]
      \includegraphics[width=\columnwidth]{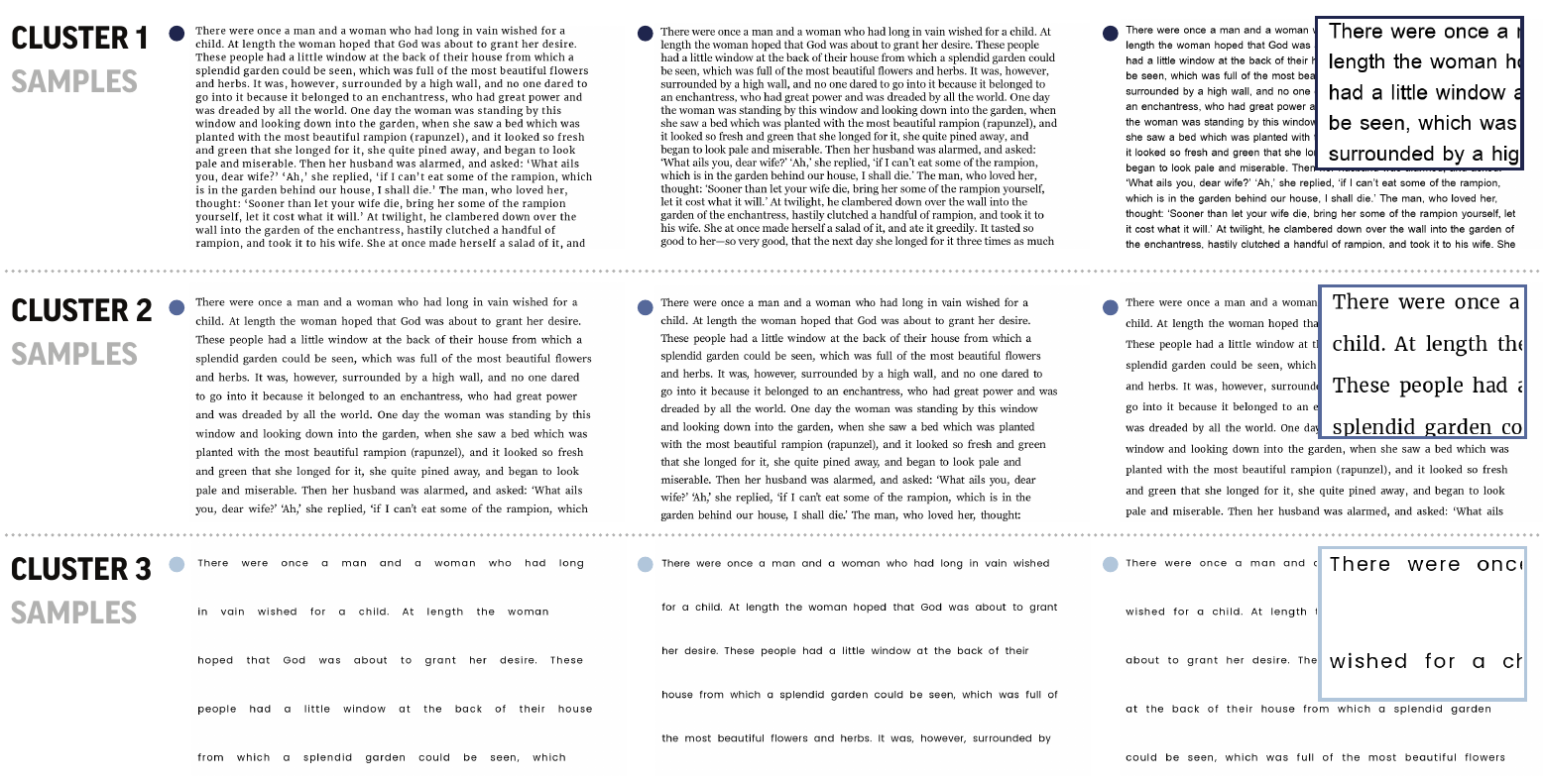}
      \caption{ \firstrevision{Three examples} of crowdworker-designed reading formats partitioned into \firstrevision{each cluster} in iteration R3. Clustering was done by a k-Means algorithm running on CNN feature vectors. Reading formats in each cluster exhibit similarities in spacing and font settings. Documents with similar fonts, such as Source Serif Pro and Times, were occasionally grouped if spacing settings were consistent.}
      \label{fig:cluster-examples}
\end{figure}

We tried a few different approaches to cluster similar reading formats together. In one approach, we used the values of the text settings (fonts, spacings) as the features for clustering. However, it is not clear how to weight these different features, as they have different effects on the final text appearance. Instead, we found that more interpretable clusters formed when we used screenshots of the reading formats. \firstrevision{We trained a convolutional neural network (CNN) on crops of the reading format screenshots to learn to group similar reading formats together (see Figure~\ref{fig:model-structure} and Appendix~\ref{appendix:cnn-details} for details). We use the trained CNN to produce feature vectors for the reading formats, which we then cluster using the k-Means algorithm (see Figure~\ref{fig:cluster-examples} for cluster examples)~\citep{vassilvitskii2006k}}. \firstrevision{To choose the number of clusters,} based on the trade-off between the number of clusters and the quality of the clusters, we used silhouette scores and knee point heuristics~\citep{rousseeuw1987silhouettes}.

Given the clusters of similar reading formats obtained in the previous step, our next step was to select a representative format from each cluster to serve as a reading theme. During the pilot study, we had asked designers to help us select cluster representatives, but this was a time-consuming process that did not achieve consensus across designers. Because we considered a reduced set of text adjustments for the main study, automatically selected cluster centroids from the k-Means algorithm ended up being good cluster representatives; when shown to designers for validation, designers indicated that they would have made similar choices.

\subsection{Stage 3: Designers refine and supplement}
\label{sec:therif-designer-supplement}

\begin{figure}[!htbp]
      \includegraphics[width=\columnwidth]{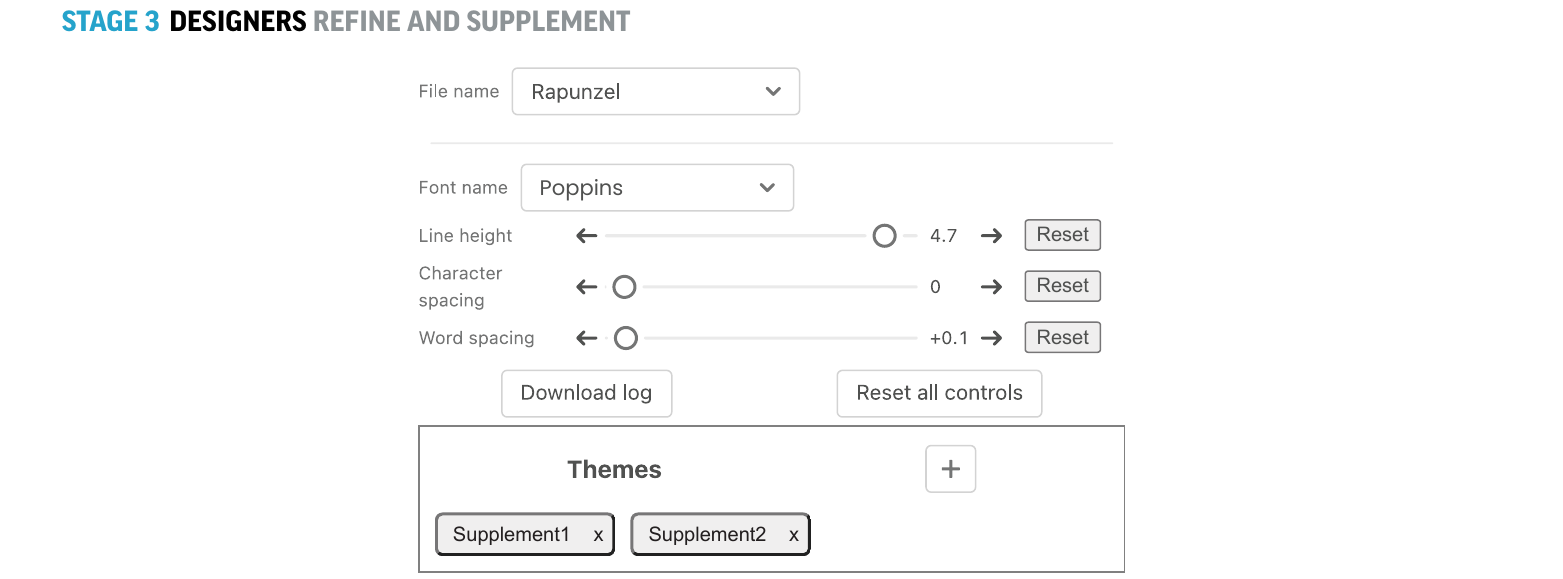}
      \caption{Text Settings Panel designers used in the main study. Compared to the interface for crowdworkers, the designer's interface does not have the theme rating panel (Figure~\ref{fig:study-interface-user-rating}) but includes the ability to store designs as supplementary reading themes. The designer's interface used during the pilot study was similar, but had the more comprehensive list of settings from Figure~\ref{fig:study-interface-pilot-text-settings}.}
      \label{fig:study-interface-designer}
\end{figure}

Document formatting, especially getting the combination of fonts and spacings to look and feel right, is an involved design task. During the pilot design sessions, D2 commented that designing reading formats that respond to different people ``compound the variables designers have to consider'', including but not limited to ``glyph characteristics and a variety of spacing features''. To incorporate design considerations into the theme feedback loop, we asked designers to evaluate the reading themes selected after automatic clustering, \firstrevision{by allowing them to view each theme and refine it further using the text settings panel.
      Designers could then} save their refinements as additional themes (Figure~\ref{fig:study-interface-designer}). \firstrevision{We used the designers from the pilot study. To keep the task tractable, one designer reviewed the themes after each iteration:} \firstrevision{D1} added 3 themes after R0, \firstrevision{D2} added 6 themes after R1, and \firstrevision{D3} added 3 themes after R2. In the last iteration (R3), we chose not to involve designers in creating additional themes or modifying the existing ones, to directly evaluate the crowdsourced themes generated by THERIF.

The themes generated from the automatic clustering supplemented by the designers' themes formed the full set of reading themes presented to participants in the next iteration of THERIF (a total of 12 themes in R1, 12 in R2, and 6 in R3). By involving both designers and new sets of crowdsourced participants in evaluating themes generated after each iteration, we continued improving the themes over time.

\subsection{Study Participants}
\label{sec:therif-participants}

We used the Prolific platform to recruit participants who used English as their first language. We recruited 200 participants for the first iteration (R0) and 100 for each subsequent iteration (R1-R3), for a total of 500 participants.  The larger number of participants in R0 was to allow the presets to deviate from the reading themes initialized from the pilot study, thus increasing the diversity of the formats in future iterations.

Dyslexia and age are important factors requiring interface adaptations when reading~\citep{rello2017good, Rello2017, Li2019, rello2013good, rello2020predicting, miniukovich2017design, Wilkins2009, Bernard2001, cai2022personalized, Calabrese2016}. In each iteration, we recruited roughly 50\% participants with dyslexia and 50\% without.\footnote{Readers with symptoms of dyslexia account for about 15-20\% of the world population~\citep{dyslexia_basics2020}. We maintained equal representation to ensure that their reading preferences were also adequately considered.}  Disclaimer: for convenience, in addition to those diagnosed with dyslexia, we will refer to participants who scored highly on a dyslexia questionnaire (see \S\ref{sec:identify-dyslexia}) as ``participants with dyslexia'' in the following paragraphs, although they may not have been formally diagnosed nor are willing to self-label as such.

We recruited participants without dyslexia equally from different age brackets: 18-25, 26-35, 36-45, 46-55, and 56-87.\footnote{Participants' age information reflects their age at the time of data export rather than the time of the study.} However, when recruiting participants with dyslexia, we were unable to recruit equally from each age group due to the limited number of participants with dyslexia on the crowdsourcing platform. Nonetheless, compared to previous readability studies~\citep{Rello2017, rello2017good, Li2019}, our study has significantly more participants with dyslexia, a more balanced representation across age groups, and overall, a larger number of participants. These participants are not intended to be a representative sample of the population in age and dyslexia, but instead to \firstrevision{explicitly include} diverse readers.

\begin{figure*}[!htbp]
      \centering
      \includegraphics[width=0.6\linewidth]{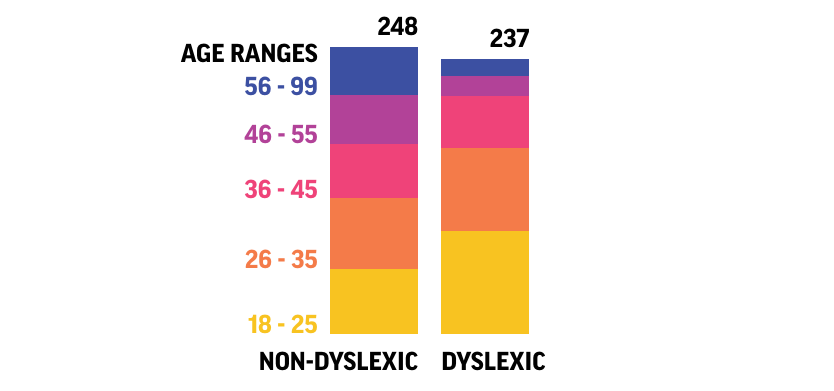}
      \caption{A total of 485 crowdworkers participated in the main study (R0-R3). We recruited equal numbers of participants with and without dyslexia and attempted to balance participants across ages by recruiting separately from each of the age buckets visualized. This was easier to do for participants without dyslexia, but participants with dyslexia were scarcer on the crowdsourcing platform. }
      \label{fig:demographics-composition}
\end{figure*}

For data quality purposes, we removed participants who (1) failed the attention check question (``Do you bike across the pacific to get to work each day?''), (2) did not complete the full study, (3) did not correctly fill in their username, or (4)~finished the study exceptionally fast (three standard deviations faster than the mean of their age group). 485 participants remained after data removal, 237 (49\%) with dyslexia and 248 (51\%) without (Figure~\ref{fig:demographics-composition}). Among the participants, 287 (59\%) are women, 196 (40\%) are men, and 2 ($<1\%$) did not answer. Studies lasted around 30 minutes on average, and participants were compensated \$13.5 hourly.

\subsubsection{Identifying participants with dyslexia}
\label{sec:identify-dyslexia}

Identifying participants with dyslexia is challenging because access to comprehensive assessments for dyslexia can be expensive~\citep{sood2018digital, bell2013professional}, and many with language difficulties go undiagnosed~\citep{germano2010comorbidity, pennington2006single, adlof2015morphosyntax, nation2004hidden}. To mitigate these challenges, we identified participants with a higher than average chance of having dyslexia using a questionnaire developed by \cite{cooper2011revised}. This questionnaire has shown effectiveness at identifying readers with dyslexia and was adopted by past readability studies~\citep{snowling2012validity, helland2011predicting, wolff2002prevalence, zorzi2012extra}. We conducted a screening study with 1,608 Prolific participants who reported having difficulty reading. Among them, we identified 397 (25\%) participants who may have dyslexia, and 237 took part in our study.

\subsection{Final themes}
\label{sec:final-themes}

\firstrevision{Each iteration of THERIF results in a set of reading themes, obtained by clustering crowdsourced text adjustments (\S\ref{sec:NN}). The cluster representatives, which are the reading themes obtained from a given iteration, are then used as input to the next iteration. After four iterations of THERIF, three clusters remained, and their corresponding cluster representatives are our final three reading themes, without any further designer inputs or adjustments. Although additional THERIF iterations could have been run, we noticed minimal changes from the third to the fourth iteration (see~\S\ref{ssec:themesconverge} and \S\ref{ssec:lesscust}). Although additional designer input and adjustments could have been made, we found no significant differences between designer-curated themes and the automatically selected cluster representatives in earlier iterations (\S\ref{sec:theme-eval-with-baseline}). In the next section, we evaluate these three themes, and offer them as a crowdsourced design contribution of this work.}

Each of the three themes represents a distinct reading format. Reading themes with larger line spacing also have correspondingly larger character and word spacing. For convenience, we name them the \textbf{COR themes}: Compact, Open, and Relaxed. \firstrevision{We visualize what text looks like when rendered in these three themes in}  Figure~\ref{fig:final-reading-themes} \firstrevision{and include CSS to reproduce the themes in Appendix~\ref{appendix:final-themes-css}}.  We additionally include the demographics of the crowdworkers whose text formats were clustered together to produce the \firstrevision{COR} reading themes~\firstrevision{(Figure~\ref{fig:final-reading-themes})}. We observed that more participants ended up in the cluster with larger spacing, although 14\% of participants preferred the most compact reading format, which is slightly below the recommended web accessibility spacing guidelines~\citep{kirkpatrick2018web, rello2012layout}. Most of the participants preferring the compact setting were 26-45, representing the portion of the population more likely to be working professionals. In our pilot study, we found that this cohort of the population was more likely to read for work and have to navigate text quickly, so a more compact format could suit these needs.  Participants older than 55 seldom chose a more compact text setting. Formats preferred by participants with dyslexia were more likely part of the reading theme with the largest spacing, supported also by previous literature~\citep{Rello2017}. Noteworthy is that this format was also preferred by a number of participants without dyslexia.

\begin{figure*}[!htbp]
      \centering
      \includegraphics[width=\textwidth]{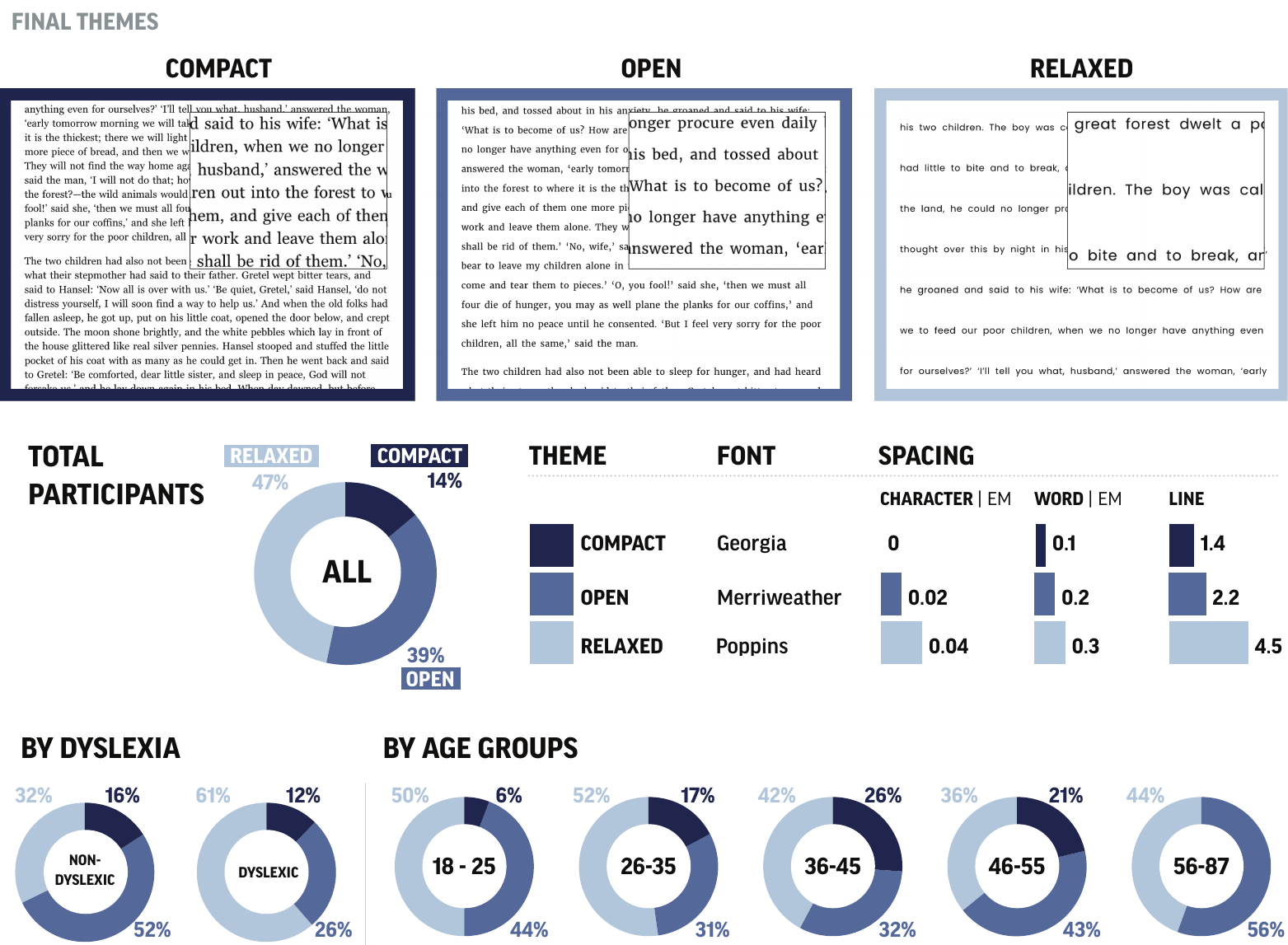}
      \caption{Our set of COR (Compact, Open, and Relaxed) reading themes are the result of the text settings of hundreds of crowdsourced participants, and thereby represent diverse reading experiences that vary in font and spacing. Note that all three spacings (character, word, and line) increase together from the Compact to the Relaxed theme. From the last full iteration of THERIF (R3), we plot the demographics of participants whose text formats were clustered to produce each of these three themes. For instance, in the top left chart, we see that almost half the participants (47\%) made text setting adjustments that corresponded to the Relaxed theme. A vast majority of participants with dyslexia had text settings that corresponded to the Relaxed theme, whereas participants without dyslexia had settings that corresponded to the Open theme. We see similar differences by age. For instance, no participants over 55 had text settings that corresponded to the Compact theme. \firstrevision{CSS values for these themes are provided in Appendix~\ref{appendix:final-themes-css}}.}
      \label{fig:final-reading-themes}
\end{figure*}

\section{Analysis and Evaluation}\label{sec:evaluation}

\firstrevision{The role of reading themes is to improve the reading experience. Because both preference and performance are part of the reading experience, in this section, we evaluate both. However, our focus on this paper is skewed towards preference (i.e., whether users like the text presets bundled with a reading theme) because the design process for themes was preference-driven to begin with. In \S\ref{ssec:difpref}-\ref{ssec:lesscust} (preference evaluation),}
we evaluate the effectiveness of THERIF at producing reading themes that are generally likable, match diverse reader preferences, and require limited further customization (i.e., the defaults are ``good enough''). \firstrevision{In \S\ref{ssec:readperf} (performance evaluation),}
we also assess the \firstrevision{COR} themes' impact on reading comfort, comprehension, and speed \firstrevision{to relate our work to prior work on font readability showing that preference and performance are often different \citep{wallace2022towards}}. 

\subsection{Different readers have different preferences}\label{ssec:difpref}

Participants had a variety of preferences when reading digitally, as demonstrated by the text settings that different readers adjusted and the final values they selected. We found some of this variation to be linked to participants' demographics. This reinforces the importance of providing multiple reading themes that fit readers' \secondrevision{diverse} preferences.

\subsubsection{There is no one-size-fits-all}

\begin{figure*}[!htbp]
      \centering
      \includegraphics[width=\linewidth]{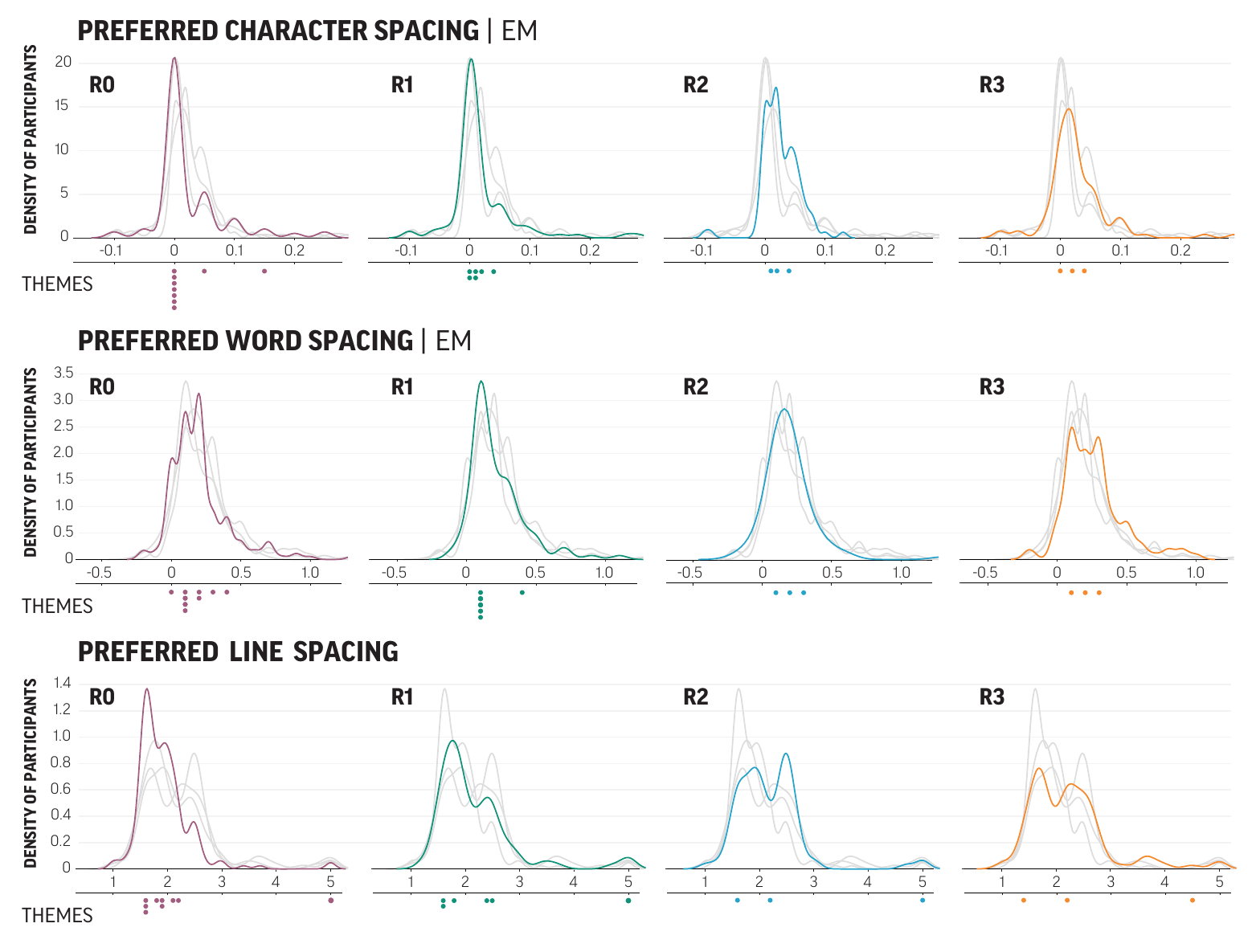}
      \caption{Distributions of the preferred text settings (histograms) and the resulting theme settings (scatters \firstrevision{below the x-axis}). With more iterations, participants explored additional text settings, as shown by the less peaked distributions. Themes' setting values start stabilizing at R2. While R0 started with 9 themes, R2 and R3 ended with 3 themes each, and the range of values that they represent (in character, word, and line spacing) nevertheless stayed similarly broad. More of the earlier themes clustered around similar values, while the later themes represent more distinct experiences. Axes drop the long tails beyond the 99th percentile, and the same smoothing factor applies to all kernel density estimations. }
      \label{fig:setting-histogram}
\end{figure*}

Distributions of preferred text settings were multi-modal; i.e., there was no single spacing or font that every participant preferred (Figures~\ref{fig:setting-histogram} and~\ref{fig:setting-histogram-fonts}).  Additionally, as the iterations of THERIF progressed, text settings became increasingly varied. This is in part because the themes that participants used as starting points diverged, allowing more of the text formatting space to be explored. Additional iterations helped consolidate the theme values, forming distinct text settings that stabilized by the fourth iteration (Figure~\ref{fig:setting-histogram}).

\subsubsection{Comparison across age groups}

When comparing the preferred text settings between age groups using a one-way ANOVA, we did not find a statistically significant difference in the preferred character spacing $(F=0.28, p=0.89)$ or line spacing $(F=2.01, p=0.09)$. However, the preferred word spacing settings differed between groups $(F=3.49, p<0.01)$. We used the TukeyHSD pairwise test for post hoc analysis. Compared to participants in the age groups of 26-35 and 46-55, those 18-25 preferred larger word spacing, averaging 0.30em compared to 0.21em and 0.20em respectively $(p<0.05, \text{Cohen's } d=0.28; p=0.03, \text{Cohen's } d=0.35)$.

\subsubsection{Comparison with and without dyslexia}

Compared to participants without dyslexia, those with dyslexia preferred larger character spacing $(t(418.76) = 2.24, p = 0.03, \text{Cohen's } d = 0.2)$, word spacing $(t(349.2) = 3.95, p < 0.01, \text{Cohen's } d = 0.36)$, and line spacing $(t(422.1) = 5.51, p < 0.01, \text{Cohen's } d = 0.5)$ (Table~\ref{tab:average-exp-dyslexia}). The significant difference in reading preferences between readers with and without dyslexia necessitates the inclusion of both groups of readers when designing reading experiences.

Participants with and without dyslexia had similarly different preferences for themes across the THERIF iterations. Themes that were downvoted by participants without dyslexia were more likely to be preferred by those with dyslexia (see Figure~\ref{fig:theme-composition}), a trend also reflected in the \firstrevision{COR} themes (Figure~\ref{fig:final-reading-themes}). Nonetheless, large overlap in preference exists.

\begin{table}[!htbp]
      \centering
      \begin{tabular}{lrrrrrr}
            \toprule
                         & \multicolumn{2}{r}{\textbf{Character Spacing} (em)} & \multicolumn{2}{r}{\textbf{Word Spacing} (em)} & \multicolumn{2}{r}{\textbf{Line Spacing}}                      \\
                         & mean                                                & std                                            & mean                                      & std  & mean & std  \\
            \midrule
            Non-Dyslexic & 0.02                                                & 0.05                                           & 0.18                                      & 0.18 & 1.93 & 0.52 \\
            Dyslexic     & 0.03                                                & 0.07                                           & 0.28                                      & 0.35 & 2.25 & 0.73 \\
            \bottomrule
      \end{tabular}
      \caption{ The average refined text settings by participants with and without dyslexia. Participants with dyslexia preferred larger character, word, and line spacing than those without dyslexia, and these differences are statistically significant. }
      \label{tab:average-exp-dyslexia}
\end{table}

\subsubsection{Comparison across iterations}
\label{sec:comparison-across-iterations}

Comparing iterations, we found that the participants' average preferred line spacing increased from 1.92 to 2.18 from R0 to R1 $(t(141.97) = 3.03, p < 0.01, \text{Cohen's } d = 0.43)$. However, no statistically significant difference in other text settings existed between iterations.  This finding indicates that the participants had consistent spacing preferences throughout iterations of the study.

\begin{figure}[!htbp]
      \includegraphics[width=\textwidth]{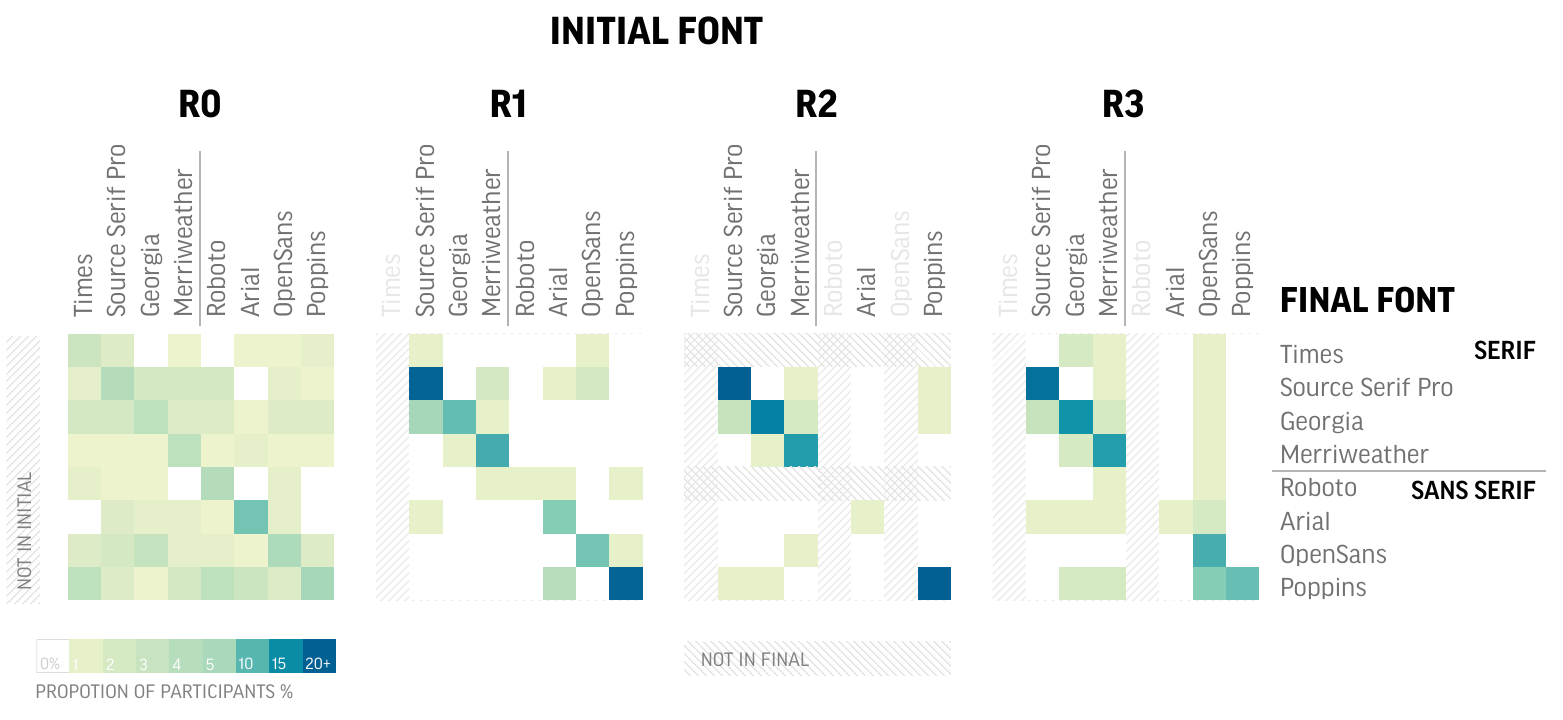}
      \caption{Participants' font refinement from the theme defaults. With additional iterations, a smaller proportion of participants refined their font choices from ones that their preferred themes were initialized with, and even fewer switched from serif to sans serif or vice versa. In comparison, a larger proportion of participants in R0 changed their font choices.}
      \label{fig:setting-histogram-fonts}
\end{figure}

Font choices did not differ significantly among participants from different age groups $(\chi^2(28, N=485)=21.6, p=0.80)$ or between participants with and without dyslexia $(\chi^2(7, N=485)=8.9, p=0.26)$. However, between iterations, the distributions of the preferred fonts varied $(\chi^2(21, N=485)=66.6, p<0.01)$. The evolving theme settings may cause such variation. In R0, where fonts are randomly paired with spacing settings from the pilot, only 36.6\% of the participants stayed with the theme's default font. However, in R1-R3, where participants selected from 12, 12, and 6 themes respectively, 76.3\%, 87.8\%, and 66.7\% respectively chose not to refine the font choice from the theme default despite the fewer number of themes available to choose from (Figure~\ref{fig:setting-histogram-fonts}).

\subsubsection{Preference for themes is diverse}
\label{sec:diverse-theme-preference}

\begin{figure*}[!htbp]
      \centering
      \includegraphics[width=\textwidth]{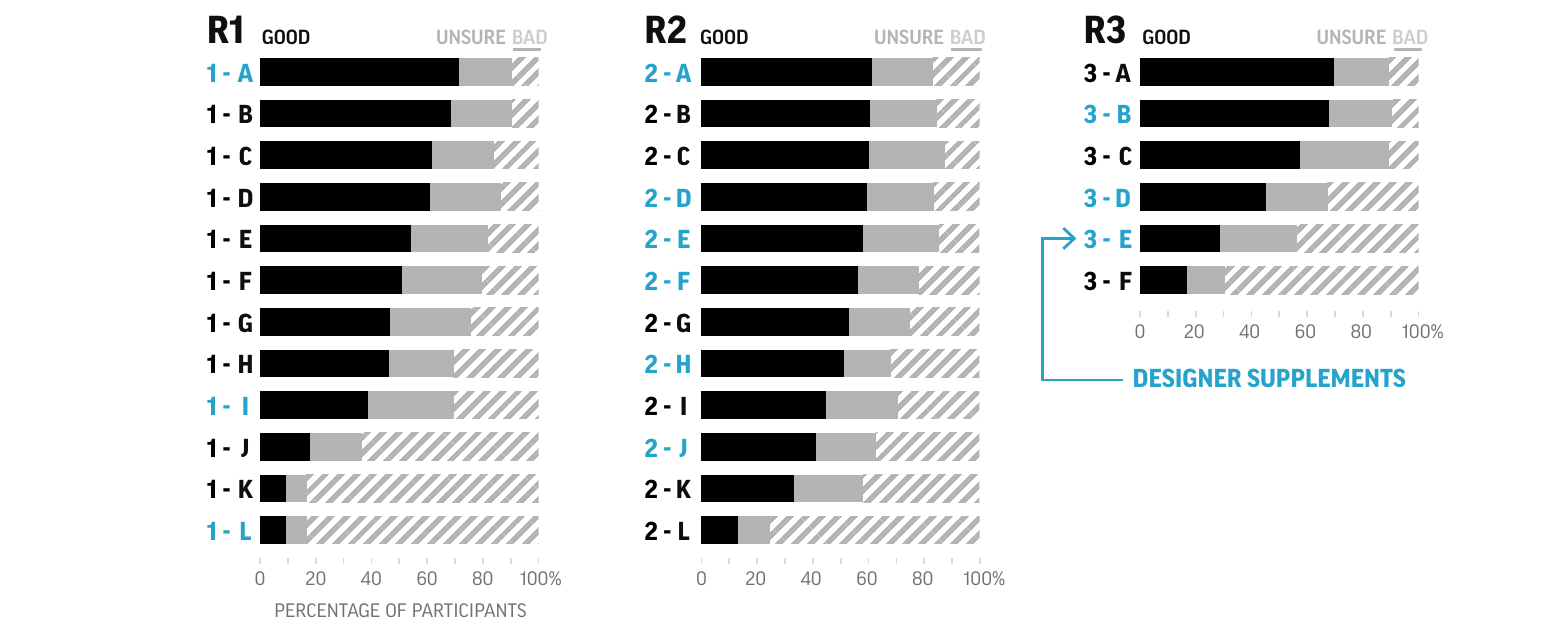}
      \caption{The number of votes received by each reading theme.  Most themes received more positive votes than neutral or negative votes.  Crowdsourced themes received similar ratings as designer-created ones. We re-labeled themes with indices for easy reference. Participants in R0 did not provide ratings because they received randomly initialized themes.}
      \label{fig:theme-preference}
\end{figure*}

After each THERIF iteration, the combination of the crowdsourced and designer-created themes formed the set of reading themes shown to new participants in the next iteration.  Both crowdsourced and designer-created themes received similarly positive responses, with most themes receiving more positive votes than unsure or negative (Figure~\ref{fig:theme-preference}). No single theme was a winner by a large margin, an unsurprising finding given participants' preference for diverse text settings (Figure~\ref{fig:setting-histogram}). As a sanity check that participants' ratings are not random, all the validation themes (\S\ref{ssec:primaryreview}) received predominantly negative votes (see supplementary material), and only 2.4\% (7 out of 294) participants from R1 - R3 indicated that they preferred a validation theme.

\subsubsection{\firstrevision{THERIF themes and designer-created themes are similarly preferred}}
\label{sec:theme-eval-with-baseline}

\firstrevision{Similar to prior literature~\citep{park2013crowd,yu2011cooks}, including a related iterative design crowdsourcing \secondrevision{pipeline}~\citep{yu2011cooks}, we compare designer creations with designs (i.e., themes) that automatically emerge from our THERIF \secondrevision{pipeline}. Specifically, participants in iterations R1-R3 of THERIF were presented with themes that were both automatically selected and manually created by designers, and we leveraged their feedback to compare these two sets of themes. Across R1-R3, the number of good, unsure, and bad votes received by designed themes and THERIF-generated themes did not differ significantly $(t(28) = 0.49, p = 0.6278, \text{Cohen's } d = 0.183; t(28) = -0.25, p = 0.8009, \text{Cohen's } d = -0.095, t(28) = -0.29, p = 0.7704, \text{Cohen's } d = -0.11)$. In R1 and R3, THERIF-generated themes received more positive votes per theme (43.8 votes in R1, 42.7 in R2, 45.3 in R3) compared to designed themes (37.7 in R1, 52.8 in R2, 44.3 in R3). When asked to select a favorite theme, similar numbers of participants selected the THERIF-generated themes (on average 7 participants per theme in R1, 8 in R2, 15 in R3) and the designed themes (on average 9 participants per theme in R1, 7 participants in R2, 17 participants in R3.}

\subsection{Themes converge over iterations to concisely represent diverse formats}\label{ssec:themesconverge}

Next, we evaluated the ability of the clusters automatically computed in the THERIF \secondrevision{pipeline} to be representative of the formats customized by participants using the provided text settings.  We first evaluated the individual clusters visually with designers. All four designers confirmed that our automatic approach effectively grouped formats together, where formats in a cluster were more similar than formats across clusters.

We also evaluated the quality of the clusters using silhouette scores, which assign higher scores to clusters of reading formats with higher \emph{intra}-cluster similarity and lower \emph{inter}-cluster similarity. In Table~\ref{tab:silhouette-score} we see that the silhouette scores for the clustering improved with additional iterations of THERIF, indicating that the formats created by participants were converging to a handful of similar experiences, that the reading themes could represent well.

Keeping the clustering criteria the same, the total number of clusters decreased over the THERIF iterations (Table~\ref{tab:silhouette-score}). A smaller number of clusters, which translates to a smaller number of themes, makes the process of selecting a comfortable reading format easier. However, we wanted to ensure that even with the smaller number of clusters, diverse text settings were still represented. Indeed, we observed that the ranges of the text settings values remained relatively stable between iterations (Figure~\ref{fig:setting-histogram}).  The settings between R2 and R3 were more similar compared to the previous iterations, suggesting convergence. With additional iterations, themes with similar text settings were grouped, leading to increasingly distinct settings. For instance, R0 and R1 saw clusters of character spacing and line spacing settings similar in value. However, they were later grouped into the same theme (Figure~\ref{fig:setting-histogram}).

\begin{table}[!htbp]
      \centering
      \begin{tabular}{lll}
            \toprule
            Iteration & Number of Clusters & Silhouette Score \\
            \midrule
            R0        & 9                  & 0.19             \\
            R1        & 6                  & 0.22             \\
            R2        & 3                  & 0.20             \\
            R3        & 3                  & 0.26             \\
            \bottomrule
      \end{tabular}
      \caption{Silhouette scores improve with additional study iterations. Silhouette scores range from -1 to 1 and measure the quality of clustering. Dense clusters that are well separated from each other achieve a score closer to 1. }
      \label{tab:silhouette-score}
\end{table}

\begin{figure*}[!htbp]
      \centering
      \includegraphics[width=\textwidth]{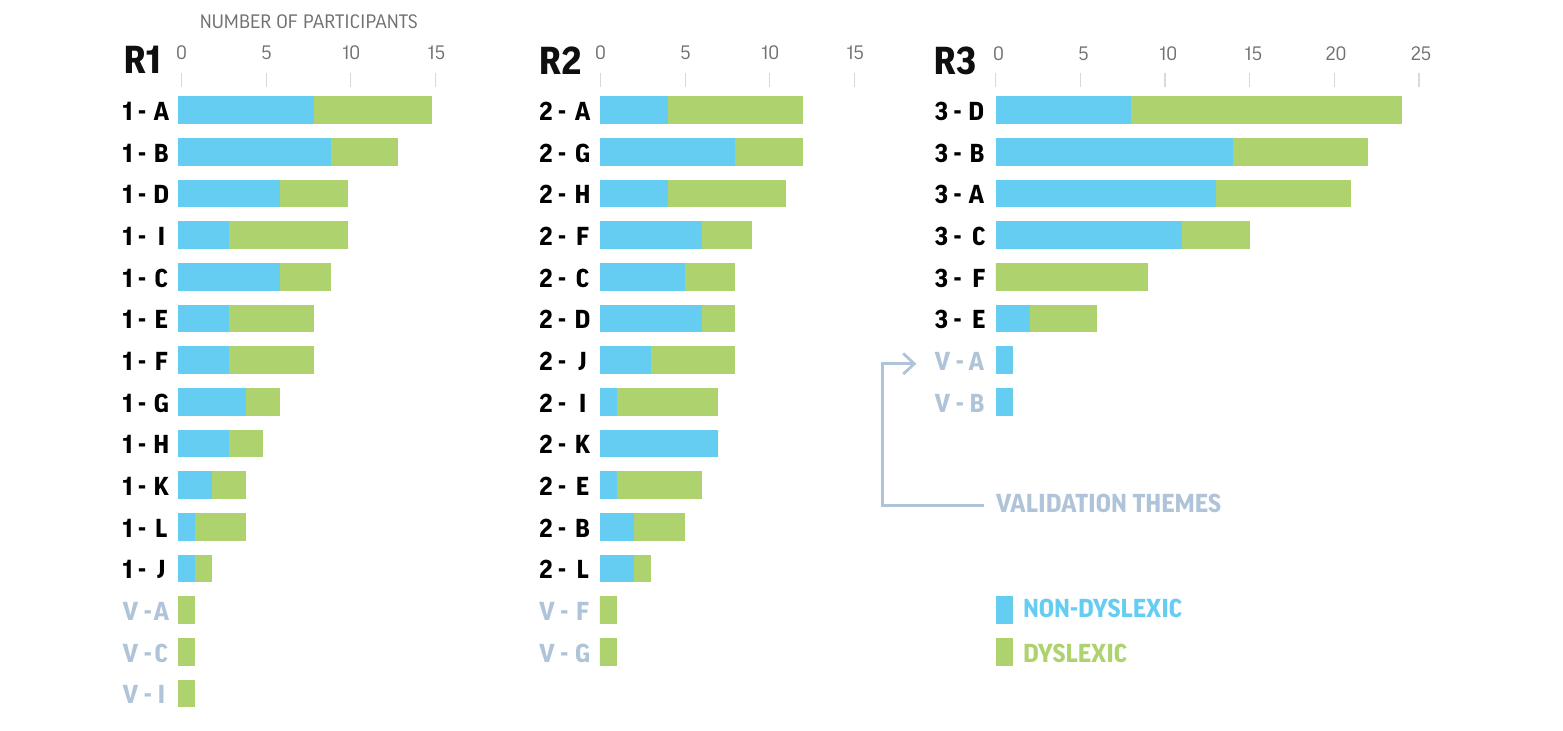}
      \caption{Participants' most preferred themes (themes labeled with indices for convenience). Note that some themes were more likely to be preferred by participants with dyslexia, and other themes by participants without dyslexia, but the considerable overlap in preferences reflects the need to design for different participants inclusively, rather than to design separate experiences. Labels starting with ``V'' refer to validation themes (poorly designed experiences). Throughout THERIF iterations R1-R3, only 7/485 participants found a validation theme preferable. Participants in R0 did not select preferred themes because they received randomly initialized themes.}
      \label{fig:theme-composition}
\end{figure*}

\subsection{Themes in later iterations require less customization}\label{ssec:lesscust}

\begin{figure*}[!htbp]
      \centering
      \includegraphics[width=\linewidth]{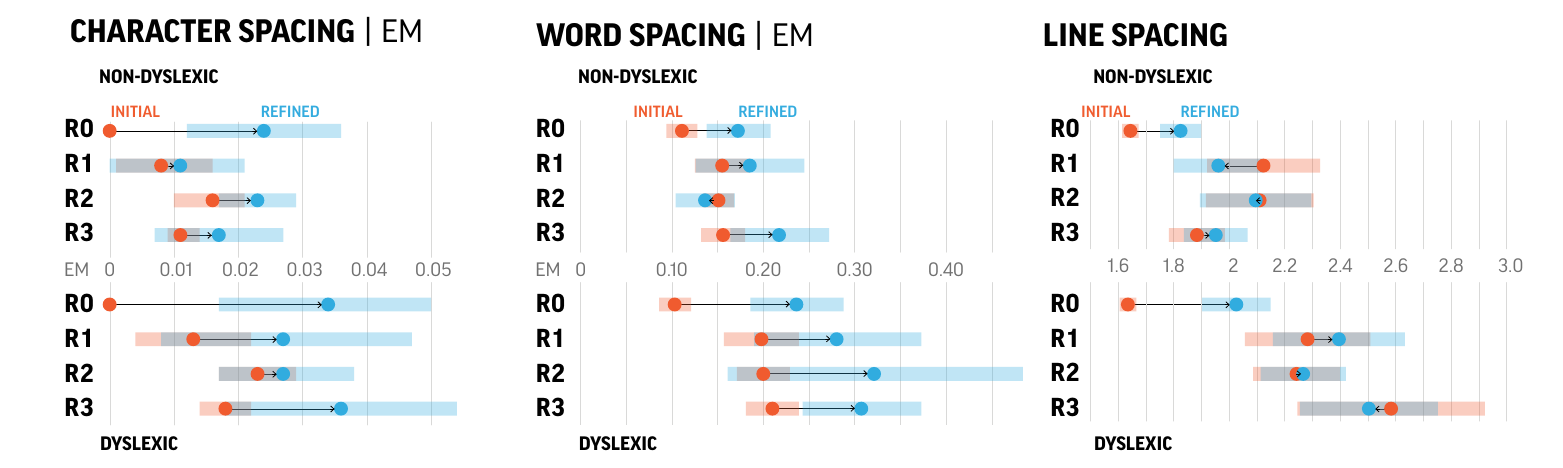}
      \caption{Differences between default reading theme settings and the subsequently refined text settings participants ended up with. For themes in the later iterations, participants deviated less from the default theme settings. 95\% confidence intervals are visualized.}
      \label{fig:changes-from-theme}
\end{figure*}

During the initial refinement iteration (R0), we observed that all participants increased character, word, and line spacing from the default theme, resulting in a set of diverse formats (Figure~\ref{fig:changes-from-theme}). Participants with dyslexia increased word spacing during R1 and R3, and participants without dyslexia increased character spacing in R2 and word spacing in R3, respectively (Table~\ref{tab:within-iter-changes}). As the THERIF iterations progressed, participants made fewer refinements to the provided themes (Figure~\ref{fig:changes-from-theme}), indicating that the defaults were ``good enough'', i.e., meeting their preferences.

\begin{table}[!htbp]
      \centering
      \begin{tabular}{lllrrrrl}
            \toprule
            Iteration & Dyslexic & Text Setting      & df & t    & p    & Cohen's d \\
            \midrule
            R0        & Yes      & Character Spacing & 90 & 4.03 & 0.00 & 0.60      \\
            R0        & Yes      & Line Spacing      & 90 & 6.15 & 0.00 & 0.89      \\
            R0        & Yes      & Word Spacing      & 90 & 5.47 & 0.00 & 0.72      \\
            R0        & No       & Character Spacing & 99 & 3.94 & 0.00 & 0.56      \\
            R0        & No       & Line Spacing      & 99 & 4.59 & 0.00 & 0.62      \\
            R0        & No       & Word Spacing      & 99 & 3.64 & 0.00 & 0.44      \\
            R1        & Yes      & Word Spacing      & 47 & 2.18 & 0.04 & 0.33      \\
            R2        & No       & Character Spacing & 48 & 4.14 & 0.00 & 0.37      \\
            R3        & Yes      & Word Spacing      & 48 & 3.42 & 0.00 & 0.54      \\
            R3        & No       & Word Spacing      & 49 & 2.58 & 0.01 & 0.41      \\
            \bottomrule
      \end{tabular}
      \caption{Refinements from the reading themes made by participants with and without dyslexia, based on paired t-tests. Participants made significant refinements in all three spacing settings from the randomly initialized reading themes shown in the initial refinement iteration. The iterations that followed saw fewer adjustments in comparison.}
      \label{tab:within-iter-changes}
\end{table}

\begin{figure}[!htbp]
      \includegraphics[width=0.6\columnwidth]{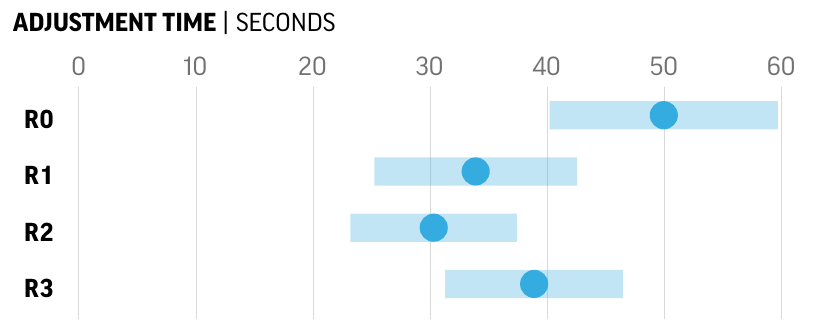}
      \caption{Average time participants spent adjusting text settings in each of the THERIF iterations. Compared to starting from a randomly initialized theme in R0, starting from a theme (R1-R3) reduced the average time participants spent on refining the text settings. Note that R3 had 50\% fewer themes than previous iterations, but the refinement time did not change significantly.}
      \label{fig:adjustment-duration}
\end{figure}

Apart from examining the text settings themselves, we also considered the time participants spent making adjustments to the default themes.  Comparing iterations, we found that participants in R1 spent significantly less time on refinement than R0 $(t(272.39) = -2.42, p = 0.02, \text{Cohen's } d = -0.26)$.\footnote{We applied Welch's t-test to compare adjustment times between R0 and R1 due to their unequal variances and sample sizes. We applied t-tests of equal variance for subsequent comparisons. Only time spent on refinement after the ``Secondary Theme Review'' step is considered adjustment time (Figure~\ref{fig:stage-one-decomposed}).}  Refinement time may be related to the number of themes presented. The fewer theme defaults available, the more likely participants will spend more time making refinements, which explains why participants in R0 with one default setting spent the most time customizing their reading format. However, despite the decreasing number of themes from 12 in R1 to 6 in R3, participants did not spend more time on refinements. No significant difference existed when comparing the refinement time between R1 and R2 or R2 and R3 (Figure~\ref{fig:adjustment-duration}), indicating that the smaller number of themes were nevertheless meeting participant preferences.

\subsection{Reading themes can improve reading performance}\label{ssec:readperf}

\firstrevision{While the focus of this work is on offering reading themes that match participants' preferences as shown in the previous evaluation sections, in this section we also consider how themes can contribute to reading performance.}

When identifying participants with dyslexia in our screening study (\S\ref{sec:identify-dyslexia}), we also surveyed them about factors they considered most important when reading digitally. Of the 1,608 respondents, 75.6\% and 71.1\% respectively considered reading comprehension and comfort important, while 34.9\% cared about speed.\footnote{This was administered as a multiple-choice question. Participants could provide additional free-form text answers; no other consistent criteria surfaced.} \firstrevision{Here we evaluate whether themes summarized from crowdsourced, preference-based designs can also} improve reading performance, \firstrevision{measured by these three objectives readers considered important}.

\subsubsection{Methods}

We conducted readability tests using an interface modeled after the one in \cite{wallace2022towards}. During the study, participants read in the three COR themes, as well as a control theme that mimics standard digital reading \firstrevision{defaults in Microsoft Word and Google Doc} (Arial font, 0em character, 0em word, 1 line spacing). \secondrevision{Consistent with prior readability studies, we choose \firstrevision{a fixed control theme to help compare themes' effect across \secondrevision{participants across age and dyslexia spectrums~\citep{Li2019, wallace2022towards, kadner2021adaptifont, wery2017effect}.}}}

\firstrevision{Participants read an 8th-grade passage in each theme, presented in a randomized order. Passages averaged 150-250 words in length, and they were split across four separate screens to capture more robust reading speed measurements while maintaining comparable numbers of words per screen~\citep{cai2022personalized, wallace2022towards}. Each passage was followed by four comprehension questions~\citep{wallace2022towards}. To quickly transition between screens, participants made a key press. To get acquainted with the interface, participants completed a warm-up study round (in a format different from the four tested themes).}

We measured a reading theme's \firstrevision{performance} using three metrics: reading comfort, comprehension, and speed. We measured the comfort score using a 5-point Likert scale answer (``not at all'' to ``extremely'') to the question ``how comfortable is the reading experience you've just seen?'', speed using WPM (Words Per Minute), and comprehension as the percentage of questions answered correctly.

\subsubsection{Study Participants}
\label{sec:effectiveness-participants}

We recruited crowdworkers on Prolific who speak English as their first language. We removed participants who (1) did not complete any given portion of the study, (2) attempted the study multiple times, or (3)~had taken part in similar reading studies or had participated in the THERIF iterations. Based on recommendations from prior work~\citep{Carver1990, Carver1992, wallace2022towards, cai2022personalized}, we removed individual reading speed measurements above 650 or below 50 to remove participants who may be skimming or not paying attention to a given screen. \firstrevision{140 participants remained after data removal, 72 with dyslexia and 68 without. Participants without dyslexia came from a balanced age group resembling those from the main study. The study lasted 30 minutes on average, and participants were compensated \$15 hourly.}

\subsubsection{Results}
\label{sec:effectiveness-results}

\paragraph{\firstrevision{Comfort, Comprehension, and Speed}}

\begin{figure*}[!htbp]
      \centering
      \includegraphics[width=\linewidth]{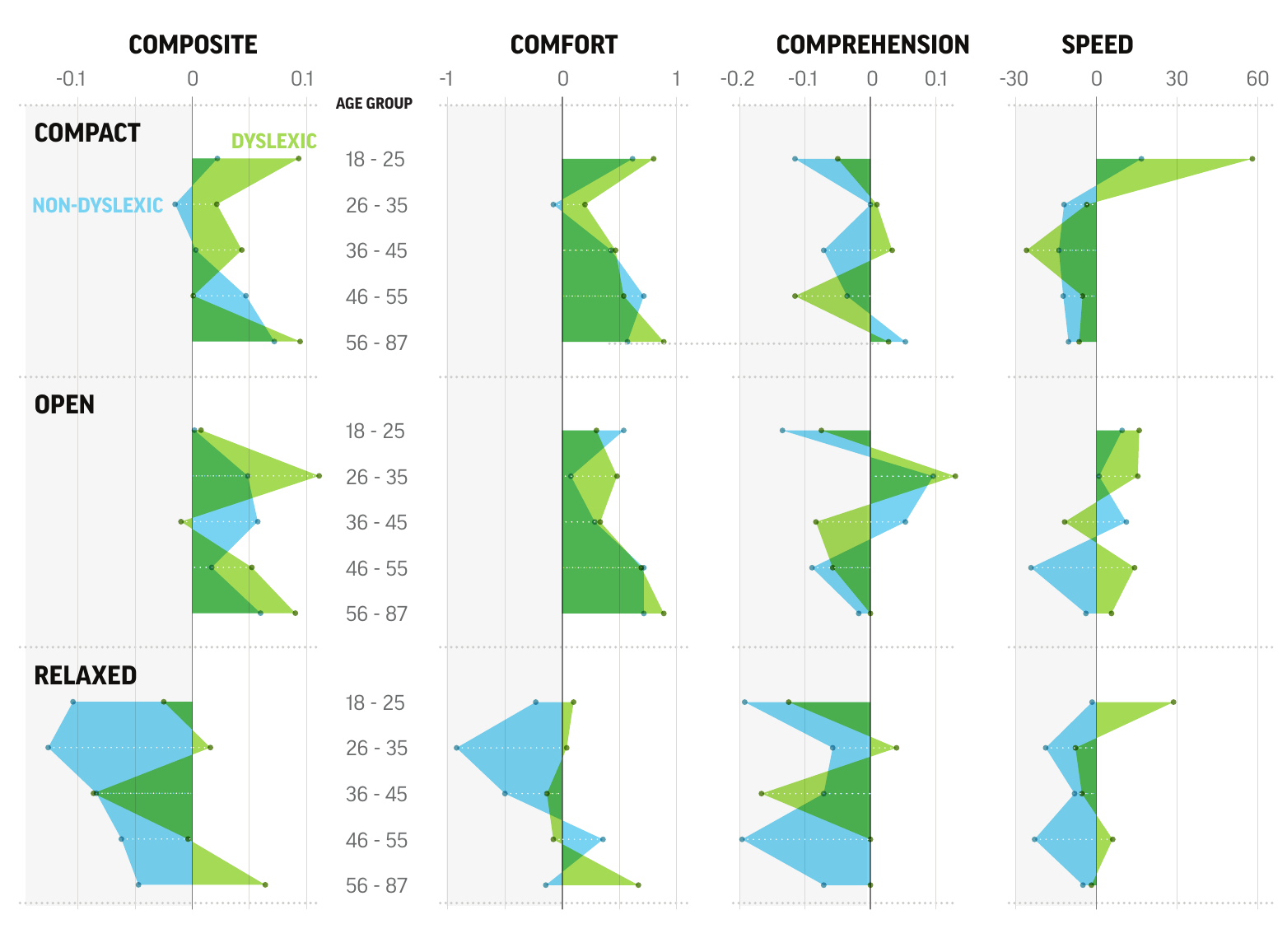}
      \caption{We compared the \firstrevision{performance} of the COR reading themes to the control theme based on objectives readers considered important (Equation~\eqref{composite-score}). The \firstrevision{first column of} charts illustrates the composite \firstrevision{performance} score, and the \firstrevision{other three columns show} individual comfort, comprehension, and speed \firstrevision{metrics}. One or more themes \firstrevision{performed better than} the control theme for participants in different groups. Participants with dyslexia benefited most from reading themes.}
      \label{fig:theme-effectiveness}
\end{figure*}

\firstrevision{ When evaluated by individual performance metrics, participants generally considered COR (Compact, Open, Relaxed) themes to be more comfortable than the control theme (Figure~\ref{fig:theme-effectiveness}). Across participants, \firstrevision{91\%} of them rated at least one of the COR themes to be at least as comfortable as the control theme, and \firstrevision{61\%} of them rated a COR theme as strictly more comfortable. Themes' effects on comprehension and speed differed by age and dyslexia. Participants with dyslexia generally read faster with the Open theme. However, the Compact theme lead to faster reading speeds for participants with dyslexia aged 18-25 (Figure~\ref{fig:theme-effectiveness}). We hypothesize that this may be due to narrow character spacing reducing saccades~\citep{Arditi1990, minakata2021effect}. No other consistent effect of reading themes on speed and comprehension was observed, as there is likely to be a speed-comprehension trade-off at play~\citep{foraker2011comprehension, wallace2021considering, reed1973speed, mackay1982problems}.}

\firstrevision{To examine whether the effect of COR themes differs by participant demographics, we constructed linear mixed effect models (LMEs) to predict each of the performance metrics with age, dyslexia, and reading theme as fixed effects, and participant ID as crossed random effects. LME results showed that age significantly affected reading speed in the Compact theme, reducing speed by 0.7 WPM per year $(t(547)=-2.130, p=0.03)$. This finding is consistent with our observation in THERIF iterations, where few older readers preferred the Compact theme~(Figure~\ref{fig:final-reading-themes}).  We did not find statistically significant variation in comfort or comprehension across age or dyslexia. See Appendix~\ref{appendix:lme} for the full results.}

\paragraph{Combining \firstrevision{performance metrics}}

\firstrevision{Evaluating reading performance with individual metrics does not account for trade-offs and interactions, such as between reading speed and comprehension~\citep{foraker2011comprehension, wallace2021considering, reed1973speed, mackay1982problems}. There is no right answer when it comes to combining the metrics, which depends on the application and scenario. In an attempt to combine the metrics into a single composite score according to what readers consider important, we use the results of our survey and convert participant votes into scalar weights to trade-off comfort, comprehension, and speed:} $\text{\firstrevision{Composite performance score}} = 42\%\times\text{Comprehension} + 39\%\times\text{Comfort} + 19\%\times\text{Speed}~\refstepcounter{equation}(\theequation)\label{composite-score}$.

\firstrevision{For each of the participant groups, at least one of the COR reading themes performed better} than the control theme (Figure~\ref{fig:theme-effectiveness}). Results differed with participants' demographics. Participants aged 18-25 performed better with the Compact reading theme, whose line spacing is larger than the control theme (1.4 instead of 1). Participants aged 26-45 performed better with the Open reading theme. Participants over 45 benefited from multiple reading themes\firstrevision{, where the Relaxed theme especially benefited older participants with dyslexia}. Generally, themes brought larger improvements to readers with dyslexia and \firstrevision{the youngest and oldest of readers in our study population.}

\section{Discussion}

In this paper, we showed that by iterating between crowd-generated text formats (through setting adjustments), automatic clustering, and design sessions, we converged on a handful of representative reading formats, which we call themes. This outcome was not guaranteed at the outset of the study, since another possible outcome could have been a growing and continually diverging set of formats. After four iterations (R0-R3) of our THERIF \secondrevision{pipeline}, multiple pieces of evidence suggested convergence: (1) The settings between iterations R2 and R3 stabilized (Figure~\ref{fig:setting-histogram}), (2) the number of unique clusters (i.e., themes) decreased across the iterations and then stopped changing (Table~\ref{tab:silhouette-score}), and (3) the adjustments participants made to the themes provided --- in terms of both absolute value (Figure~\ref{fig:changes-from-theme}) and time (Figure~\ref{fig:adjustment-duration}) decreased. In other words, participants were able to select out of the defaults provided and were satisfied enough with the initial theme settings to not have to make further changes. As iterations progressed, themes that were poorly rated earlier fell out of consideration, even though we did not explicitly filter by likability. Instead, this happened naturally as a byproduct of the automatic clustering of the most common formats study participants selected.

Importantly, each iteration of our THERIF \secondrevision{pipeline} involved a new group of participants, meaning that the themes generated from the text settings of the prior's iterations participants remained representative. If we had used the same participants for all four iterations, we would be limited to claiming that our themes converged to represent the preferences of a particular group; instead, we can be more confident in claiming that our themes represent the preferences of a population \secondrevision{across age and dyslexia spectrums}.

By recruiting participants with different reading and learning abilities, and a variety of ages, we end up converging on a set of themes that meet diverse readers' \secondrevision{preferences}. While some segments of the population tend to get more benefits from the reading themes (e.g., older participants and those who score higher on the dyslexia questionnaire both preferred, and performed better in, the themes with larger spacing), the themes are intended to be universal. Another way to look at these results is that while the effects of text settings on reading may be governed by some demographic characteristics, reading preferences and performances are highly variable across individuals, as found also by prior work~\citep{beier2021readability, wallace2022towards, cai2022personalized, Chatrangsan2019, zhu2021effects, Rello2016, McKoon2016, korinth2020wider}.

\subsection{Learnings}
\label{sec:discussion-learnings}

The learnings from this work are based on evidence that has come together from multiple sources, over the duration of our experiments: (1) the pilot studies and initial design iterations (\S\ref{sec:pilot}); (2) four iterations of the THERIF \secondrevision{pipeline} (\S\ref{sec:therif}), which included (i) text settings manually adjusted by crowdsourced participants, (ii) clusters automatically computed from all the crowdsourced formats, and (iii) design refinements along with the additional comments provided by designers; and (4) a performance evaluation of the \firstrevision{COR} themes according to multiple criteria (\S\ref{sec:evaluation}). We took care to avoid having participants participate in our studies more than once, since we had many iterations and evaluations to complete; the cumulative learnings come from the participation of \firstrevision{896} total crowdworkers\footnote{Out of \firstrevision{896} crowdsourced participants, 271 took part in the pilot study (\S\ref{sec:pilot}), 485 in the main study (\S\ref{sec:therif-participants}), and \firstrevision{140 in the reading performance} study (\S\ref{sec:effectiveness-participants}).} aged 18-87 and the involvement of 4 designers with 8 - 27 years of design and typography experience.

\paragraph{Larger character, word, and line spacing go hand-in-hand}

Considering the text setting values in the \firstrevision{COR} themes, spacing monotonically increased from one theme to the next, in line with common typographical considerations and past research~\citep{highsmith2020inside, reynolds2004you}, and similar to patterns observed in spacing combinations obtained from the pilot study (\S\ref{sec:pilotthemes}). This was not enforced but fell out naturally from the THERIF iterations. The variation in spacing between the themes is what led to our naming convention: Compact, Open, and Relaxed (Figure~\ref{fig:final-reading-themes}). This result mirrors the importance designers and participants placed on varying spacing. For instance, when refining reading themes, designers emphasized their focus on varying spacing to achieve different visual ``textures'' and ``density'' in order to support different readers. Participants favoring a ``compact reading experience'' cited the need to ``fit as much information on a page as possible''. In contrast, participants opting for larger line spacing did so to ``focus on each line of the text'' when reading. Some user interfaces, such as Gmail, already allow users to adjust spacing settings to achieve different ``information density''~\citep{gmail}. However, to our knowledge, this is the first time such a rigorous process has been undertaken to arrive at a set of spacing presets based on diverse reader \secondrevision{preferences}.

\paragraph{Larger spacing benefits readers with dyslexia}

Participants that scored high on the dyslexia questionnaire preferred larger character, word, and line spacing (Table~\ref{tab:average-exp-dyslexia}), reinforcing the finding that the three types of spacing go hand in hand. Simultaneously, a sizable proportion of the participants without dyslexia had a similar preference for increased spacing. Spacing preferences vary even among participants with dyslexia. The Relaxed theme has a line spacing of 4.5, larger than the average 2.25 setting manually selected by participants with dyslexia (Table~\ref{tab:average-exp-dyslexia}), and corresponds to a subset of participants preferring line spacing of 4-5 (Figure~\ref{fig:setting-histogram}). This setting is also larger than prior recommendations, although previous work focused on performance while we focused on reader preference~\citep{Rello2017, british2012dyslexia}. In our \firstrevision{reading performance} study, we found that older participants with dyslexia especially benefited from the Relaxed theme due to its positive effect on reading comfort and comprehension~(\S\ref{sec:effectiveness-results}).

\paragraph{Font choice varied with spacing}

Theme fonts varied with the spacing settings (Figure~\ref{fig:final-reading-themes}). Both Compact and Open themes are paired with serif fonts. While the tighter spacing in these themes increases information density, the need to accommodate serifs introduces additional horizontal spacing beyond existing text settings to support glyph legibility~\citep{arditi2005serifs}. On the other hand, Poppins, a sans serif font, was paired with the Relaxed theme. Previous research showed that sans serif fonts can better support readers with low vision and those with dyslexia~\citep{russell2007legibility, rello2016effect}. Similarly, \cite{korinth2020wider} found large character spacing to support language learners. \firstrevision{Previous work has shown that a font's impact on reading is driven by characteristics such as x-height, character width, stroke contrast, etc.~\citep{cai2022personalized}. Therefore, the recommended fonts in our COR themes may be replaceable by others with similar attributes (in case there are application, device, or branding constraints).}

\paragraph{Themes agree with accessibility guidelines and previous work}

WCAG recommends character, word, and line spacing values of 0.12em, 0.16em, and 1.5 respectively~\citep{kirkpatrick2018web}.  The settings of the Compact theme are slightly below the recommendation but are nonetheless above modern browsers' default settings~\citep{network31css}. Both Open and Relaxed themes conform to the recommendations for word and line spacing. Interestingly, participants in our study preferred narrower character spacing than recommended by WCAG.

The fonts selected for our reading themes have been shown by previous work to correspond to participant preference or good reading performance. \cite{cai2022personalized} used the same eight fonts we started our THERIF iterations with and reported that Merriweather (Open theme) was the most preferred font by participants, Georgia (Compact theme) resulted in the largest improvement in reading performance, and Poppins (Relaxed theme) led to speed improvements, especially for language learners~\citep{cai2022personalized, korinth2020wider}.\footnote{ \cite{cai2022personalized} did not normalize font sizes.}

\paragraph{Participants overlapped in their theme preferences}

THERIF did not produce separate reading themes tailored to readers with dyslexia and those without (Figure~\ref{fig:theme-composition}). Instead, all three \firstrevision{COR} reading themes reflected both cohorts' preferences. The overlap in theme preferences shows that creating separate designs based solely on different demographics (e.g., ``a font for people with dyslexia'') may be shortsighted. Unsurprisingly, our preliminary experiments were unable to predict preferred text settings based on demographic information alone (see supplementary material for results). Previous work similarly showed that individual abilities vary on a spectrum~\citep{cooper2011revised, snowling2012validity}, and a separate design for a specific cohort may not meet the full range of user needs. This leads to the realization that designing a wide enough range of experiences can meet the needs of many without needing to assign labels to people.

\subsection{Limitations and future work}

We have presented a \secondrevision{pipeline} for designing reading themes that meet the \secondrevision{diverse preferences of} English-speaking readers by iterating on crowdsourced designs, designer refinements, and an automated clustering algorithm.  Below, we discuss several limitations of our current study and provide recommendations for future research directions.

\paragraph{Participants}

Our efforts were focused on adult readers (ages 18-87) who speak English as their first language. We were not able to recruit enough participants with dyslexia that were in the older age groups (46 and above) due to the limited number of such participants on the crowdsourcing platform we used. Additionally, compared to a professional diagnosis, the dyslexia questionnaire used may fail to differentiate between participants with dyslexia and those with ADHD because they exhibit similar reading difficulties. We did not explicitly recruit participants with other conditions, such as those with low vision. Future research can expand recruitment efforts and deploy more comprehensive questionnaires to include readers that were not represented in our studies and better understand their preferences and needs.

\paragraph{Reading platform}

Our reading themes were developed in a desktop reading setting. Future work may consider generalizing the THERIF \secondrevision{pipeline} to other platforms, such as mobile devices, tablets, and e-readers, or even beyond digital reading to printed material. \firstrevision{In Appendix~\ref{appendix:othercontexts}, we report the results of a survey showing participants' willingness to use the themes in other contexts.} Some of these platforms may be better suited to specific audiences, like children in the classroom or readers in under-served communities where mobile devices may be the default reading device.

\paragraph{Reading contexts and tasks}

Previous studies reported that preferred reading formats may differ by context, such as time of day, type of reading, etc. We recruited a large number of participants to capture not only \secondrevision{a variety of} demographics but also \secondrevision{diverse} reading contexts.  Future work may consider explicitly matching reading formats to specific contexts or tasks. Whereas we focused on general reading of page-length texts, other formats may be more suitable for glanceable reading~\citep{Sawyer2020}, long-form reading~\citep{MohamadAli2013, Sawyer2020, srivastava2021mitigating}, and reading on complex backgrounds (e.g., in video captioning and AR environments)~\citep{hall2004impact, beier2021readability, Bednarski2013, rello2017good, Sawyer2020, beier2021readability}, or in the context of document elements like figures and tables~\citep{beier2021readability}.

\paragraph{Typographical considerations}

We normalized the sizes of our study fonts to obtain comparable x-height (\S\ref{sec:norm-fontsize}), an important factor influencing readability~\citep{cai2022personalized, wallace2022towards, Wilkins2009, rolo2021type, Sheedy2005, highsmith2020inside}. Previous literature and typographers interviewed for this study believed that such normalization leads to more perceptually similar reading experiences across fonts~\citep{wallace2022towards}. However, such normalization does not account for inconsistencies in character width, a factor influencing readability~\citep{minakata2021effect, beier2021increased}, and may lead to slight variations in spacing settings, which generally correlate with font sizes rather than x-heights~\citep{network31css}. Our CNN-based approach for clustering reading formats makes the THERIF \secondrevision{pipeline} robust against the effect of font normalization and changes in CSS units. Nonetheless, future work seeking finer control over the reading interface may consider tracing the glyphs' vector path data for more precise text measurements~\citep{cai2022personalized}, or conducting perceptual user studies for more accurate normalization~\citep{wallace2022towards}. Additionally, adjustments in character and word spacing may obscure the typographer's design considerations, such as kerning and ligature. Future work may consider preserving these properties when exploring the effect of font and spacing on reading.

\paragraph{Extensions to the THERIF \secondrevision{pipeline}}

The THERIF \secondrevision{pipeline} can be extended in a number of ways, and to suit other applications.  For instance, increasing the number of iterations over a longer period may help adapt reading themes to changing reader preferences. Similar to \cite{yu2011cooks}, our evidence suggests that the iterations can also proceed without explicit designer input, if it is not available, \firstrevision{particularly because we did not find differences in participants' preferences for automatically-selected and designed themes}. On the other hand, designers' involvement can help steer reading formats towards certain parts of the space, for instance, if there are any specific design needs~\citep{park2013crowd}.  Further, because clustering is performed automatically using machine learning algorithms, THERIF can scale to any number of participants and iterations.

\section{Conclusion}

The digital reading applications available today occasionally offer readers custom control over certain text settings like font, size, or spacing. Prior readability research has moreover demonstrated the benefits of personalization on reading performance itself, as measured by reading speed and comprehension~\citep{cai2022personalized, wallace2022towards, Chatrangsan2019, zhu2021effects, Rello2016, McKoon2016}. However, for the casual reader, adjusting these text settings can be cumbersome (\S\ref{sec:personalizing-text-settings-is-challenging}). For instance, adjustments to character or word spacing can change the look and feel of the text, which may in turn require compensatory adjustments to the other spacing or font parameters. Instead of leaving this text tuning process in the hands of the casual reader, we propose providing readers with reading themes: preset combinations of fonts and spacings. To arrive at reading themes that would cater to readers \secondrevision{across age and dyslexia spectrums}, we used an iterative feedback loop, involving crowdworkers, automatic clustering, and designer input, \firstrevision{similar to relevant \secondrevision{pipeline} in~\citep{nickerson2008spatial, yu2011cooks, park2013crowd, gulley2001patterns, resnick2009scratch}}. We demonstrated that our \secondrevision{pipeline}, called THERIF, was successful in producing themes that met the preferences of diverse readers, bringing them to their preferred reading formats faster.

Four iterations of our THERIF \secondrevision{pipeline} converged on three themes with increasing character, word, and line spacing when moving from Compact to Open and Relaxed themes. Font also varied between the themes, with serif (Georgia and Merriweather) fonts selected for the first two themes, and a sans serif font (Poppins) selected for the Relaxed theme. In our studies, participants over 55 preferred the two themes with the larger spacings, and a significant proportion of participants with high scores on the dyslexia questionnaire preferred the Relaxed theme. Nevertheless, all three COR themes catered to \secondrevision{readers' diverse preferences}. The THERIF iterations were run on a total of 485 participants, and the earlier pilot studies featured another set of 271 participants, all of whom contributed to the learnings that shaped the \firstrevision{COR} themes. A survey of 1,608 participants showed that comfort and comprehension outweigh speed as the key measures for reading \firstrevision{performance}, and a group of \firstrevision{140} participants achieved better reading outcomes with reading themes developed by THERIF than a control theme similar to a default web browser format.

While professional designers participated in our iterative feedback loop, we did not find that their inputs significantly affected the outcome of our study. In particular, themes that were tweaked by designers were equally likely to be chosen by crowdsourced participants as the automatically suggested themes\firstrevision{~(\S\ref{sec:theme-eval-with-baseline})}. The THERIF \secondrevision{pipeline} is extendable to future iterations, and our evidence points to the fact that it can be run without further designer intervention.

In our studies, participant demographics were correlated with text setting preferences. On the one hand, this points to a possible future of automatically suggesting reading themes to readers, similar to the individualized font predictions in \cite{cai2022personalized}. On the other hand, there is no one-to-one mapping between participant characteristics and reading formats. So unlike the approach of designing for a subset of the population (e.g., dyslexic fonts, or speed reading tools), our \secondrevision{pipeline} has led to themes that cater to \secondrevision{diverse} readers' preferences, and would not require readers to be explicitly labeled or to label themselves. Our work brings us a step closer to allowing every reader, struggling or proficient, young or old, to read comfortably. Where most reading today occurs on digital surfaces, text that caters to individual reader needs should be the rule, not the exception. This is where customization and inclusivity go hand-in-hand.

\section*{Acknowledgements}

We thank Surabhi Bhargava, Jose Echevarria, Narendra nath Joshi, Joy Kim, Qisheng Li, Mauli Pandey, Shaun Wallace, and Yi-le Zhang for their valuable feedback and edits. We thank Tim Brown, Natalie Dye, Astha Kabra for sharing their perspectives on design and typography. 

\newpage
\begin{appendices}

{\color{\firstmarkupcolor}\makeatletter\let\default@color\current@color\makeatother

\section{Text settings in pilot and main studies}

See Table~\ref{tab:reading-control-comparison}.

\begin{table}[!htbp]
    \centering
    \begin{tabular}{lcc}
          \toprule
          Text Settings      & Pilot & Main \\
          \midrule
          Character Spacing  & +     & +    \\
          Word Spacing       & +     & +    \\
          Line Spacing       & +     & +    \\
          Font Name          & +     & +    \\
          Font Size          & +     & --    \\
          Paragraph Indent   & +     & --    \\
          Paragraph Spacing  & +     & --    \\
          Column Width       & +     & --    \\
          Text Alignment     & +     & --    \\
          Color and Contrast & +     & --    \\
          Dark Mode          & +     & --    \\
          \bottomrule
    \end{tabular}
    \caption{ The main study includes a subset of the text settings used in the pilot study. Settings marked ``+'' could be adjusted by the participants in the main study, and those marked ``-'' were fixed after the pilot study. The pilot study helped us identify which settings lead to systematically different preferences across participants. Based on participant and designer feedback, we removed settings unrelated to readability (e.g., paragraph spacing) or that vary considerably across reading contexts (e.g., dark mode). During the main study, all fonts were shown at the same x-height as 17px Times, paragraphs had no indent, paragraphs each had 1em spacing before and after, and the column width was 6in. All texts were left-aligned, in black, over a white background. }
    \label{tab:reading-control-comparison}
\end{table}

\section{Cluster crowdsourced reading formats with CNN and K-Means}
\label{appendix:cnn-details}

We trained a convolutional neural network (CNN) on crops of reading format to group similar reading formats together in a self-supervised way. We reproduced screenshots of participants' reading formats from their refinements log files (similar to the examples in Figure~\ref{fig:teaser}; see supplementary material for real samples). We then used random crops of the screenshots to train a CNN to predict the source (participant ID) of each crop, i.e., a self-supervised training approach (Figure~\ref{fig:model-structure}). This ensured that the CNN model learned to associate crops from the same reading formats, which would allow it to later group similar formats together. Diversity in participant preferences made self-supervised training viable. The model was trained on data from R0, where 191 participants created 174 (91\%) distinct reading formats. We experimented with the crops and selected a size (128px per side, equivalent to 3.36 deg of visual angle) that captured multiple lines of text at varied spacings. The final model achieved an accuracy of 79\% on the test set. We then used the feature vectors from the penultimate layer of the trained CNN for clustering.

We ran the k-Means algorithm on feature vectors of each crop to group together similar formats designed by different participants (Figure~\ref{fig:cluster-examples})~\citep{vassilvitskii2006k}. We used 1000 random crops from each reading format to thoroughly represent the format and increase clustering robustness. We followed the knee point heuristics to select the appropriate number of clusters using the algorithm from~\cite{satopaa2011finding}, with a smoothing factor of 2. We did not consider results from more than 20 clusters, as it would be unrealistic for real-world readers to choose from this many reading themes.

\section{Comparison of text settings in THERIF}
\label{appendix:theirf-comparisons}

See Table~\ref{tab:summarize-statistics}.

\begin{table}[htbp!]
    \centering
    \begin{tabular}{llll}
          \toprule
                            & \textbf{Age Group}                & \textbf{Dyslexia}                          & \textbf{Study Iteration}    \\
          \midrule
          Character Spacing &                                   & w/ dyslexia\textgreater w/o dyslexia &                             \\
          Word Spacing      & 18-25\textgreater 26-35 and 46-55 & w/ dyslexia\textgreater w/o dyslexia &                             \\
          Line Spacing      &                                   & w/ dyslexia\textgreater w/o dyslexia & R1\textgreater R1           \\
          Font              &                                   &                                            & Differed between iterations \\
          \bottomrule
    \end{tabular}
    \caption{A summary of statistical tests results on the different in text settings by age group, dyslexia, and study iteration.}
    \label{tab:summarize-statistics}
\end{table}

\section{COR Themes in CSS}
\label{appendix:final-themes-css}

See Table~\ref{tab:final-themes-css}.

\begin{table}[htbp!]
    \centering
    \begin{tabular}{llll}
          \toprule
                                & \textbf{Compact} & \textbf{Open} & \textbf{Relaxed} \\
          \midrule
          characterSpacing (em) & 0                & 0.02          & 0.04             \\
          wordSpacing (em)      & 0.1              & 0.2           & 0.3              \\
          lineHeight            & 1.4              & 2.2           & 4.5              \\
          fontName              & Georgia          & Merriweather  & Poppins          \\
          fontSize (px)         & 15.8             & 15.8          & 14.1             \\
          \bottomrule
    \end{tabular}
    \caption{The three final themes' CSS values.}
    \label{tab:final-themes-css}
\end{table}

\section{Performance Consistency between 8th and 12th-grade Passages}

A week later after the study with 8th-grade passages, we re-recruited 25 out of 140 participants to read four 12th-grade passages in the same four themes. 12th-grade passages averaged 250-350 in length~\citep{wallace2022towards}, and they were split across six separate screens followed by four comprehension questions. For the 25 participants that completed both studies (with 8th and 12th-grade passages), we evaluated whether their reading performance improved \emph{consistently} when reading with the same theme both times. In 88\% of cases, the theme that improved the reading speed of a participant in the first study (with 8th-grade passages) also improved the reading speed of the same participant in the second study (with 12th-grade passages) relative to the control theme. In 52\% of cases comprehension scores were consistent --- i.e., the same theme led to comprehension improvements relative to the control in both studies. In 72\% of the cases the same theme was rated as more comfortable to read in compared to the control in both studies. While this is a limited study with only 25 repeat participants, these initial results provide some evidence that the benefits participants receive from the themes that work best for them are consistent over time and reading levels (at least comparing 8th to 12th-grade reading).

\section{Results of Linear Mixed-Effect Models}
\label{appendix:lme}

We constructed linear mixed effect models (LMEs) to predict each of the performance metrics with age, dyslexia, and reading theme as fixed effects, and participant ID as crossed random effects. A participant-level random effect creates separate intercepts per participant to reflect their varying reading performance. We included interaction terms between age, reading theme, and dyslexia to understand how the effect of themes on reading \firstrevision{performance} may differ by participant's age and dyslexia. See Tables~\ref{tab:lme-comfort_rating_response}, \ref{tab:lme-score}, and \ref{tab:lme-mean_wpm}.

\begin{table}[htbp!]
    \centering
    \begin{tabular}{lrrrrrr}
          \toprule
          {}                     & Coef.  & Std.Err. & z      & $P>|z|$ & [0.025 & 0.975] \\
          \midrule
          Intercept              & 3.134  & 0.313    & 10.009 & 0.000 & 2.520  & 3.747  \\
          theme=compact          & 0.196  & 0.378    & 0.519  & 0.604 & -0.544 & 0.936  \\
          theme=open             & 0.135  & 0.378    & 0.359  & 0.720 & -0.605 & 0.876  \\
          theme=relaxed          & -0.674 & 0.378    & -1.783 & 0.075 & -1.414 & 0.067  \\
          dyslexic               & -0.231 & 0.191    & -1.214 & 0.225 & -0.605 & 0.142  \\
          dyslexic:theme=compact & 0.047  & 0.230    & 0.204  & 0.839 & -0.404 & 0.498  \\
          dyslexic:theme=open    & 0.065  & 0.230    & 0.281  & 0.778 & -0.386 & 0.515  \\
          dyslexic:theme=relaxed & 0.374  & 0.230    & 1.626  & 0.104 & -0.077 & 0.825  \\
          age                    & -0.005 & 0.007    & -0.735 & 0.462 & -0.018 & 0.008  \\
          age:theme=compact      & 0.006  & 0.008    & 0.764  & 0.445 & -0.010 & 0.022  \\
          age:theme=open         & 0.008  & 0.008    & 0.985  & 0.325 & -0.008 & 0.024  \\
          age:theme=relaxed      & 0.009  & 0.008    & 1.158  & 0.247 & -0.007 & 0.025  \\
          Group Var              & 0.343  & 0.083    &        &       &        &        \\
          \bottomrule
    \end{tabular}
    \caption{Results of a linear mixed-effect model predicting the participant's comfort rating. Group variable is the participant ID uniquely identifying each study participant and is incorporated as random effects.}
    \label{tab:lme-comfort_rating_response}
\end{table}

\begin{table}[htbp!]
    \centering
    \begin{tabular}{lrrrrrr}
          \toprule
          {}                     & Coef.  & Std.Err. & z      & $P>|z|$ & [0.025 & 0.975] \\
          \midrule
          Intercept              & 0.792  & 0.066    & 11.920 & 0.000 & 0.662  & 0.922  \\
          theme=compact          & -0.104 & 0.094    & -1.104 & 0.270 & -0.288 & 0.080  \\
          theme=open             & 0.017  & 0.094    & 0.178  & 0.858 & -0.167 & 0.201  \\
          theme=relaxed          & -0.180 & 0.094    & -1.917 & 0.055 & -0.364 & 0.004  \\
          dyslexic               & -0.048 & 0.040    & -1.176 & 0.240 & -0.127 & 0.032  \\
          dyslexic:theme=compact & 0.024  & 0.057    & 0.414  & 0.679 & -0.088 & 0.136  \\
          dyslexic:theme=open    & 0.023  & 0.057    & 0.403  & 0.687 & -0.089 & 0.135  \\
          dyslexic:theme=relaxed & 0.083  & 0.057    & 1.459  & 0.145 & -0.029 & 0.196  \\
          age                    & -0.000 & 0.001    & -0.151 & 0.880 & -0.003 & 0.003  \\
          age:theme=compact      & 0.002  & 0.002    & 0.835  & 0.404 & -0.002 & 0.006  \\
          age:theme=open         & -0.001 & 0.002    & -0.415 & 0.678 & -0.005 & 0.003  \\
          age:theme=relaxed      & 0.001  & 0.002    & 0.739  & 0.460 & -0.002 & 0.005  \\
          Group Var              & 0.000  & 0.009    &        &       &        &        \\
          \bottomrule
    \end{tabular}
    \caption{Results of a linear mixed-effect model predicting the participant's comprehension score. Group variable is the participant ID uniquely identifying each study participant and is incorporated as random effects.}
    \label{tab:lme-score}
\end{table}

\begin{table}[htbp!]
    \centering
    \begin{tabular}{lrrrrrr}
          \toprule
          {}                     & Coef.    & Std.Err. & z      & $P>|z|$ & [0.025  & 0.975]  \\
          \midrule
          Intercept              & 292.151  & 24.655   & 11.850 & 0.000 & 243.828 & 340.474 \\
          theme=compact          & 22.585   & 15.251   & 1.481  & 0.139 & -7.306  & 52.476  \\
          theme=open             & 14.072   & 15.251   & 0.923  & 0.356 & -15.819 & 43.963  \\
          theme=relaxed          & -6.322   & 15.251   & -0.415 & 0.678 & -36.213 & 23.569  \\
          dyslexic               & -8.326   & 15.009   & -0.555 & 0.579 & -37.742 & 21.091  \\
          dyslexic:theme=compact & 4.433    & 9.284    & 0.477  & 0.633 & -13.763 & 22.629  \\
          dyslexic:theme=open    & 8.852    & 9.284    & 0.954  & 0.340 & -9.344  & 27.048  \\
          dyslexic:theme=relaxed & 12.036   & 9.284    & 1.296  & 0.195 & -6.160  & 30.232  \\
          age                    & -0.921   & 0.531    & -1.736 & 0.083 & -1.962  & 0.119   \\
          age:theme=compact      & -0.699   & 0.328    & -2.130 & 0.033 & -1.343  & -0.056  \\
          age:theme=open         & -0.372   & 0.328    & -1.134 & 0.257 & -1.016  & 0.271   \\
          age:theme=relaxed      & -0.120   & 0.328    & -0.367 & 0.714 & -0.764  & 0.523   \\
          Group Var              & 6313.625 & 24.140   &        &       &         &         \\
          \bottomrule
    \end{tabular}
    \caption{Results of a linear mixed-effect model predicting the participant's reading speed. Group variable is the participant ID uniquely identifying each study participant and is incorporated as random effects.}
    \label{tab:lme-mean_wpm}
\end{table}

\section{Themes could generalize to other devices and contexts}\label{appendix:othercontexts}

When asked what kind of reading they would use their chosen themes for, 37.1\% of participants expressed willingness to use their preferred reading theme on a variety of platforms and content. Separately, 23.8\% and 17.3\% of participants expressed interest in using themes for reading on a computer or reading long passages, two use cases included in our study setup (Table~\ref{tab:theme-future-application}).

\begin{table}[!htbp]
    \centering
    \begin{tabular}{lr}
          \toprule
          Application               & Percentage \\
          \midrule
          All of the above          & 37.1       \\
          Reading on computer       & 23.8       \\
          Reading long passage      & 17.3       \\
          Reading on mobile devices & 6.8        \\
          Reading on tablet         & 6.5        \\
          Reading email             & 4.8        \\
          Reading short passage     & 2.7        \\
          Reading social media post & 0.3        \\
          Others                    & 0.7        \\
          \bottomrule
    \end{tabular}
    \caption{When asked how they would use the reading themes beyond the scope of this study, the majority of the participants indicated a willingness to continue using themes. ``All of the above'' indicates all other pre-specified options. }
    \label{tab:theme-future-application}
\end{table}

}
\clearpage
\newpage

\end{appendices}


\bibliographystyle{ACM-Reference-Format}
\bibliography{main}

\end{document}